\documentclass[]{aa}
\usepackage{natbib}
\usepackage{hyperref}
\usepackage{graphicx}
\usepackage{amsmath}
\usepackage{amssymb}
\usepackage{txfonts}
\usepackage{color}

\begin{document}

\title{Accounting for Stochastic Fluctuations when 
Analysing Integrated Light of Star Clusters. I: First Systematics}

\author{Morgan Fouesneau\inst{1} \and Ariane Lan\c{c}on\inst{1} }

\offprints{M. Fouesneau \email{morgan.fouesneau@astro.unistra.fr}  }

\institute{
 Observatoire Astronomique de Strasbourg,
 Universit\'e de Strasbourg \& CNRS (UMR 7550), Strasbourg, France
}

\date{\textbf{Accepted April 23, 2010 }}

\authorrunning{Fouesneau \& Lan\c{c}on}
\titlerunning{Estimating star cluster properties}

\abstract
{Star clusters are studied widely both as benchmarks for stellar evolution
models and in their own right. Cluster age distributions and mass distributions
within galaxies are probes of star formation histories, and of cluster formation
and disruption processes. The vast majority of clusters in the Universe is
small, and it is well known that the integrated fluxes and colours of all but
the most massive ones have broad probability distributions, due to small numbers
of bright stars.}
{This paper goes beyond the description of predicted probability distributions,
and presents results of the analysis of cluster energy distributions in an
explicitly stochastic context.}
{The method developed is Bayesian. It provides posterior probability
distributions in  the age-mass-extinction space, using multi-wavelength
photometric observations and a large collection of Monte-Carlo simulations of
clusters of finite stellar masses. The main priors are the assumed intrinsic
distributions of current mass and current age for clusters in a galaxy. Both
UBVI and UBVIK data sets are considered, and the study conducted in this paper
is restricted to the solar metallicity.}
{We first use the collection of simulations to reassess and explain errors
arising from the use of standard analysis methods, which are based on continuous
population synthesis models: systematic errors on ages and random errors on
masses are large, while systematic errors on masses tend to be smaller. The
age-mass distributions obtained after analysis of a synthetic sample are very
similar to those found for real galaxies in the literature. The Bayesian
approach on the other hand, is very successful in recovering the input ages and
masses over ages ranging between 20 Myr and 1.5 Gyr, with only limited
systematics that we explain.}
{Taking stochasticity into account is important, more important for instance
than the choice of adding or removing near-IR data in many cases.  We found no
immediately obvious reason to reject priors inspired by previous (standard)
analyses of cluster populations in galaxies, i.e. cluster distributions that
scale with mass as $M^{-2}$ and are uniform on a logarithmic age scale.}

	  
\maketitle

\section{Introduction} 

In the early decades of astrophysics, star clusters have been our main key to
the understanding of stellar evolution. While clusters continue to provide
precious constraints on stellar physics, they are today studied in their own
right and as tracers of the histories of galaxies. It has become clear that a
significant fraction of star formation occurs in clusters, and that events such
as interacting galaxies can trigger their formation
\citep{Harris1991,Meurer1995, Barton2000, DiMatteo2007}. Questions have been
raised regarding the IMF in clusters in various environments, about the
systematic trends in their colour distributions, about their lifetimes as
gravitationally bound objects and about the initial and current cluster mass
functions. 

Resolved observations of individual stars remain the most precise way of
investigating the nature of clusters and will be possible out to distances of 10
Mpc with future extremely large telescopes. However measurements of the
integrated light of unresolved star clusters reach far beyond this scale already
today, and will remain the path of choice for the studies of large samples.

All our studies of individual clusters  and of cluster populations in galaxies
rest on our ability to estimate their current ages, masses and metallicities,
while accounting for extinction. The standard method of analysis of integrated
cluster light is based on the direct comparison of the observed colours with
predictions from {\em continuous population synthesis models}.  These models
predict fluxes with the assumption that each mass bin along the stellar mass
function (SMF) is populated according to the average value given by this SMF.
Studies based on continuous population synthesis models have led to results that
have a large impact on today's description of cluster ``demographics". For
instance, it is now usually admitted that the current cluster mass function
decreases with mass as a power law with an index close to $-2$ \citep{
Zhang1999, Bik2003, Boutloukos2003} and the debate on the cluster survival rate
also rests on distributions obtained using continuous models
\citep{Vesperini1998,Fall2001,Lada2003,Rafelski2005}.  The continuous approach
has been coupled with statistical data analysis, for instance to provide the
impression that including near-IR photometry (K band) solves the age-metallicity
degeneracy for clusters \citep{Goudfrooij2001, Puzia2002, Anders2004,
Bridzius2008}. {Still in the context of continuous population synthesis,
\citet{Fernandes2010} followed by \citet{Delgado2010} developed a Bayesian
analysis of the integrated spectra of star clusters.}

The continuous population synthesis models are strictly valid only in the limit
of a stellar population containing an infinite number of stars. Real clusters,
however, count a finite number of stars. Furthermore most of the light is
provided by a very small number of bright stars, in particular in the near-IR.
The so-called {\em stochastic fluctuations} in the integrated photometric
properties are the result of the random presence of these luminous stars.  Some
of these can be quantified using selected information provided by continuous
population synthesis models \citep[e.g.][]{Lancon2000, Cervino2002, Cervino2004,
Cervino2006}, but others require the use of {\em discrete population synthesis
models} \citep{Barbaro1977, Girardi1993, Bruzual2002,Deveikis2008,Popescu2009,
Piskunov2009}. The predicted luminosity and colour distributions depend strongly
on the total mass (or star number) in the cluster, and can be far from Gaussian
even when the total mass exceeds $10^5$~M$_{\odot}$.  The most probable colours
are offset from those predicted by continuous population synthesis when masses
are below $10^4$~M$_{\odot}$, because the single most luminous star in such
clusters will be more often on the main sequence than in the red giant phases of
evolution. Attempts to describe the colour distributions analytically have made
progress \citep[e.g.][]{Cervino2006}, but are not yet easily applicable.

The present piece of work is based on discrete population synthesis.  For the
first time, we use the discrete models not only to predict colour distributions
but to {\em analyse} the energy distributions of clusters.  We present a
Bayesian approach to the probabilistic determination of age, mass and
extinction, based on a large library of Monte-Carlo simulations of clusters.
{This method is a close analog to the one introduced by
\citet{Kauffmann2003} for the study of star formation histories in the Sloan
Digital Sky Survey.  However the variety of observable properties has completely
different origins in both contexts: stochasticity at a given age, mass and
metallicity plays a predominant role here, while different star formation
histories provide all the diversity in the model collections used for galaxy
studies. } We compare determinations based on the Bayesian approach with
traditional estimates, thus providing a new insight into systematic effects and
their consequences.  In this first paper, we focus on data sets consisting of
either UBVI or UBVIK photometry. Future work will extent to other pass-bands and
the addition of the metallicity dimension.

\section{Synthetic populations}

The analysis of cluster colours must be based on synthetic spectra that
explicitly account for the random fluctuations due to small numbers of bright
stars.

We have constructed large catalogs of synthetic clusters using Monte-Carlo (MC)
methods to populate the Stellar Mass Function (SMF) with a  finite number of
stars. For the purposes of this paper, all synthetic clusters host simple
stellar populations (SSP), i.e. their stars are coeval and have a common initial
composition.  The synthetic clusters are generated with a discrete population
synthesis code we derived from {\sc P\'egase} \citep{Fioc1997}.  Stellar
evolution at solar metallicity is modeled with the evolutionary tracks of
\citet{Bressan1993}.  The input stellar spectra are based on the library of
\cite{Lejeune1998}, as we will be interested mostly in broad band photometry
here. The SMF is taken from \cite{Kroupa2001}. It extends from $0.1$ to
$120$~M$_\odot$.  Nebular emission (lines and continuum) is included in the
calculated spectra of young objects under the assumption that no ionizing photon
escapes.  When extinction corrections are considered,  they are based on the
standard law of \citet{Cardelli1989}.

Our current collection of MC-clusters contains two types of catalogs. 

The first set of catalogs contains collections of clusters with equal numbers of
stars. In these (earlier) catalogs, the cluster ages take $69$ values that are
distributed on an approximate logarithmic scale between $1$~Myr and $20$~Gyr.
Metallicity is solar (Z$=0.02$). These catalogs are available for clusters of
$10^3$, $3.10^3$, $6.10^3$, $10^4$, $3.10^4$, $6.10^4$,$10^5$ stars. They
contain a total of $69,000$ clusters ($1000$ for each of the $69$ time steps).

Most of the results in this paper are based on a second catalog, which consists
of $1.5.10^6$ clusters with ages, taking $309$ values, distributed between
$1$~Myr and $20$~Gyr, with masses above about $500$~M$_{\odot}$ and with
metallicities Z=$0.008$, Z=$0.02$ and Z=$0.05$.  We will only discuss solar
metallicity here.  The distribution of ages is flat on a logarithmic scale above
$20$~Myr (younger ages are currently under-represented).  The number of stars in
a given cluster is drawn randomly from a power law distribution with index $-2$.
As a result, the mass distribution of the clusters in the sample fall off
approximately as M$^{-2}$ (note that M is the current stellar mass of the
cluster, not its initial mass).  These age and mass distributions were chosen as
a possible representation of real distributions in galaxies \citep{Fall2009},
although we caution that empirical determinations in the current literature are
based on non-stochastic studies.  At this point, the $N^{-2}$ distribution
simply has the convenient feature that it includes many small clusters, those
for which the distributions of predicted properties are most complex. It would
be a significantly larger computational challenge to produce a collection with a
flatter distribution and a similar number of small clusters.  On the other hand,
the current collection includes only a small number of very massive clusters.
Although the properties of the latter are more well-behaved, we consider the
catalog incomplete above $2.10^4$~M$_{\odot}$ and focus on smaller clusters for
the time being.

\begin{figure*}
	\includegraphics[width=\textwidth]{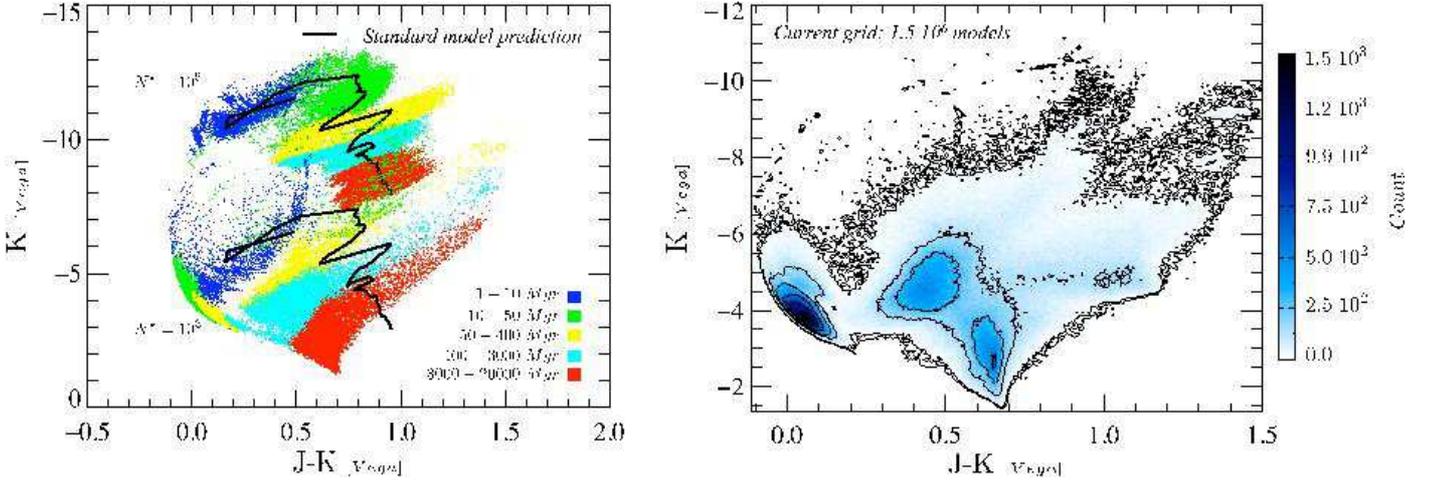}
	\caption{Stochastic properties of star clusters at solar metallicity.
	On the left panel, the dots represent photometric properties of
	individual clusters containing $10^3$ and $10^5$ stars each.  The solid
	lines show the corresponding age sequences from ``standard'' predictions
	(Figure inspired by \citet{Bruzual2002}).  On the right panel, we show
	the corresponding density distribution of models in our main catalog,
	constructed assuming a power law with a index -2 for the cluster mass
	function. {Note that the general aspect of this and subsequent model
	density maps will not change if the total number of clusters is
	increased, as required to extend the present study to higher cluster
	masses.}
	\label{fig:diags1}
	}
\end{figure*}

The differences between the properties of discrete populations and the standard
predictions from continuous populations synthesis are generally larger than
standard observational errors.  This becomes particularly true in the
near-infrared bands, and for young and intermediate ages. It requires clusters
with several $10^6 \rm{M}_\odot$ to narrow down the stochastic fluctuations to
$5$\% in the K-band \citep{Lancon2008}.  Figure~\ref{fig:diags1} illustrates the
scatter of observable properties associated with some of our synthetic cluster
catalogs, and allows direct comparison with standard predictions from continuous
synthesis. On the left panel, the lower luminosity clusters all contain $10^3$
stars (with the assumed SMF, their mass distribution is peaked around
$500$~M$_{\odot}$), while the higher luminosity clusters contain $10^5$ stars.
It is clear that the distributions are highly mass-dependent.  Most small mass
clusters have no post-main sequence star, because the average number of such
stars (given by continuous population synthesis) is lower than one. The model
density map in the right panel represents our main catalog. Due to the $N^{-2}$
distribution of star numbers and to stellar lifetimes in various evolutionary
phases, islands of high model density are present. Distributions in other
colour-colour and colour-magnitude diagrams can be found in later sections.

\section{Analysis methods}

The above synthetic populations can be used to analyse photometric observations
of clusters.  The properties we seek to estimate in this paper are the cluster
ages and masses. We wish to account for the fact that observed colours can be
affected by unknown amounts of  extinction. Accounting for uncertainties in the
metallicity is postponed.

In this section, we describe the Bayesian method developed for the analysis.
Two other analysis methods are briefly described for comparison. One is a simple
best-$\chi^2$ fit to all the data in the synthetic cluster catalog, the other is
the usual estimate based on continuous population synthesis predictions.  The
comparison between results obtained with the three methods provides important
insights into systematic effects.

\subsection{Standard estimates: the ``infinite limit"}

The standard procedure implicitly considers the mass distribution of the stars
in a cluster to be continuous. It has been applied widely, but as already
mentioned, it is strictly valid only for populations of infinite mass. It
becomes a reasonable approximation for clusters with masses above an
age-dependent limit of about $5.10^5$~M$_{\odot}$.  We include this method
mainly to quantify the errors produced when it is applied to analyse the light
of clusters of lower masses.

In families of spectra predicted with continuous models, mass is a simple
scaling factor that applies to all fluxes. Therefore model spectra are
frequently published scaled to a total population mass of $1$~M$_{\odot}$.
Spectra are described only by age and extinction (and metallicity).

Assuming observational errors are Gaussian and independent, the most likely
continuous model for a given set of broad-band fluxes is a minimum of the
following function:
\begin{eqnarray}
\chi^2_\infty = \sum_k \frac{\left( Y_k - M.Y_{k,\tilde{M_{\infty}}}
\right)^2}{\sigma_k^2},
\label{eq:chi2_std}
\end{eqnarray}
where $Y=\{Y_k\}_{k\in [1..n]}$ are the available data and the $\sigma_k$ the
corresponding observational uncertainties.  $Y_{k,\tilde{M}_{\infty}}$ are the
fluxes predicted for the model $\tilde{M}_{\infty}$ (total mass $1$~M$_{\odot}$)
and $M$ is the mass required to optimize the fit with this model. 

We allow for extinction by looping through positive values of $A_V$ and
repeating the optimization procedure. 

\subsection{Bayesian Estimates}

As opposed to its continuous analogue, this approach explicitly accounts for the
discrete and stochastic nature of the stellar populations.  The cost in
computation time is rather high, as the collections of synthetic clusters
explored must be large (see Sect.~\ref{sec:cpu}).

As in all Bayesian approaches, the results are stated in probabilistic terms,
and they depend on {\em a priori} probability distributions of some model
parameters. In our case, the most probable ages and masses for a cluster, given
a set of photometric observations, will depend on the age distribution and mass
distribution of the synthetic clusters in the model catalog (the statement can
be extended to include extinction and metallicity).

We assume observational errors are Gaussian with known standard deviations (a
preliminary study of \citet{Lancon2009} used boxcar functions for the error
distributions).  An intrinsic model property $X$, for given photometric values
$Y$, has the probability:
\begin{eqnarray}
	\mathcal{P}(X|Y) \propto \mathcal{P}(Y|X) \times \mathcal{P}(X).
\label{eq:genRelation}
\end{eqnarray}
Since errors are assumed to be Gaussian, the probability $\mathcal{P}(Y|X)$ can
be expressed using the usual $\chi^2$ statistic:
\begin{eqnarray}
	\mathcal{P}(Y|X)\propto e^{-\chi^2/2}.
\label{eq:chi2inRelation}
\end{eqnarray}
Then, the probability distribution of an intrinsic property, such as the age or
mass of an individual cluster, is given by the following relation. The
probability for property $X$ to be located in an interval $[x_1,x_2]$, given
photometric measurements $Y=\{Y_k\}_{k\in[1..n]}$ with uncertainties $\sigma_k$,
is:
\begin{eqnarray}
        \mathcal{P}(&X&\in\ [x_1,x_2]\ |\ Y) = \alpha \nonumber \\
	& \times & \sum_{\tilde{M}_i\,/\, X(\tilde{M_i})\in[x_1,x_2]} 
		\mathcal{P}(\tilde{M_i})  ~ \prod_k \frac{1}{\sqrt{2\pi\sigma_k^2}}
		~ e^{  -\frac{\left( Y_k - Y_{k,\tilde{M}_i}
		\right)^2}{2\sigma_k^2} }.
\label{eq:proba}
\end{eqnarray}
In the above, $\alpha$ is the normalization constant (the value of which we
calculate). The sum extends over all models that have an adequate value of $X$.
$\mathcal{P}(\tilde{M}_i)$ is the probability assigned to an individual model.
Unlike the continuous models, the mass is an intrinsic parameter of each
modelled population.  As a first step, we consider all models in our main
catalog equally probable, and therefore we inherit the distributions of age and
mass used to construct the catalog.  When we vary extinction, we consider flat
probability distributions for $A_V$ between two boundaries.  Through the factors
$\mathcal{P}(\tilde{M}_i)$, this expression of the probability can embody any
prior mass or age distribution. 

An analog of Eq.~\ref{eq:proba} can be written for joint probability
distributions, for instance for age, mass and extinction. The most probable ages
and masses given in this paper are the age and mass coordinates of the maximum
of the joint distribution. Error bars can be given by examining all models above
a given probability threshold.

In our main catalog, as we mentioned before, the prior mass function is a power
law with an index $-2$, and log(age) is distributed uniformly (above $50$~Myr).
Clearly, it will be necessary to investigate the effects of these assumptions
quantitatively, by varying them within reasonable limits.  Due to the above
described catalog mass completeness limit, we postpone this study to a future
work.

\subsection{Single best fit}

As a simple first step towards accounting for stochastic fluctuations, one may
look at the one model in the catalog that minimizes the standard $\chi^2$
function 
\begin{eqnarray}
	\chi^2_i = \sum_k \frac{\left( Y_k - Y_{k,\tilde{M}_i}
	\right)^2}{\sigma_k^2}.
\label{eq:chi2_fit}
\end{eqnarray}

\section{Analysis of UBVIK photometry}
\label{sec:UBVIKphot}

As a first study, we compare the results from the three methods above in the
joint analysis of U, B, V, I and K band fluxes.  The analysis is applied to the
synthetic absolute magnitudes of a subsample of clusters  from our main catalog.
The presentation is twofold: in Sect.~\ref{sec:UBVIK_nonoise_noAv} and
\ref{sec:UBVIK_nonoise_Av} the synthetic magnitudes are used unchanged, while in
Sect.~\ref{sec:UBVIK_noise} noise is added before the analysis is performed.

\subsection{Input sample description}
\label{sec:Input_sample}

\begin{figure}
        \includegraphics[width=8.8cm]{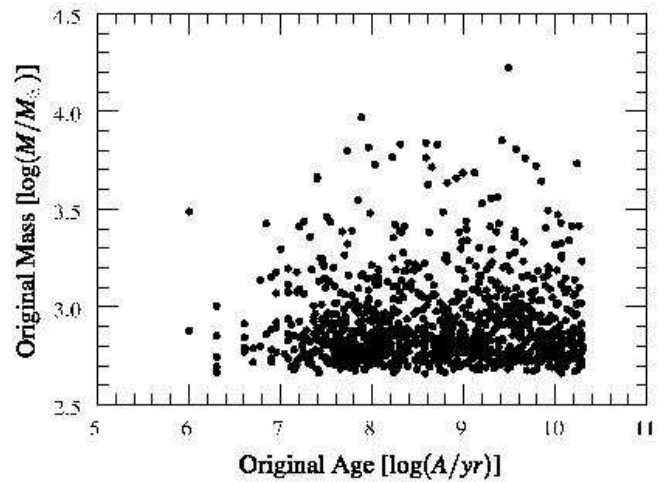}
	\caption{Model sample from our main MC catalog. 1000 models were
	randomly extracted from the catalog in order to identify systematics.
	Selected clusters are low mass populations and cover a broad range of
	ages. Fluxes are not reddened and calibrated as absolute fluxes,
	corresponding to a $10$~pc distance.
	\label{fig:UBVIK_nonoise_sample}
	}
\end{figure}

The comparison of mass and age estimates is presented for clusters with
relatively small masses, where our catalog is most complete and where the
effects of stochasticity are most important.
Figure~\ref{fig:UBVIK_nonoise_sample} shows the mass-age distribution of our
input sample. It contains 1000 models randomly selected from our main MC
catalog. It covers a relatively small mass range, but a broad range of ages
(though mostly above $10$~Myr).  The model colours are not reddened {(unless
otherwise stated)} and the model cluster distances are set to $10$ pc -- the
observable properties associated with the models are absolute magnitudes or
fluxes.  Note that the selected sample reflects the intrinsic prior mass and age
distributions of the catalog. 

\subsection{Estimates from photometry without noise}
\label{sec:UBVIK_nonoise_noAv}

{The estimates of age and mass obtained from the synthetic data with the
standard method and with the stochastic Bayesian method are shown in
Fig.~\ref{fig:UBVIK_nonoise_noAv}. Estimates from the single best fit method are
not shown, as this method returns the input model itself when no noise is added
to the synthetic data.  The standard deviations in the above equations
(\ref{eq:chi2_std}), (\ref{eq:proba}) and (\ref{eq:chi2_fit}) are set
arbitrarily to 0.05 magnitudes.  In this first experiment, the absence of
extinction, is considered to be known to the ``observer''. This could be
constrained for instance with emission line measurements or with estimations of
the total gas content of the host galaxy.  The alternative experiment, where the
extinction is a free parameter of the analysis is described in
Sect.~\ref{sec:UBVIK_nonoise_Av}.}

\subsubsection{Biases in the estimates from the standard method}
\begin{figure*}
        \includegraphics[width=8.8cm]{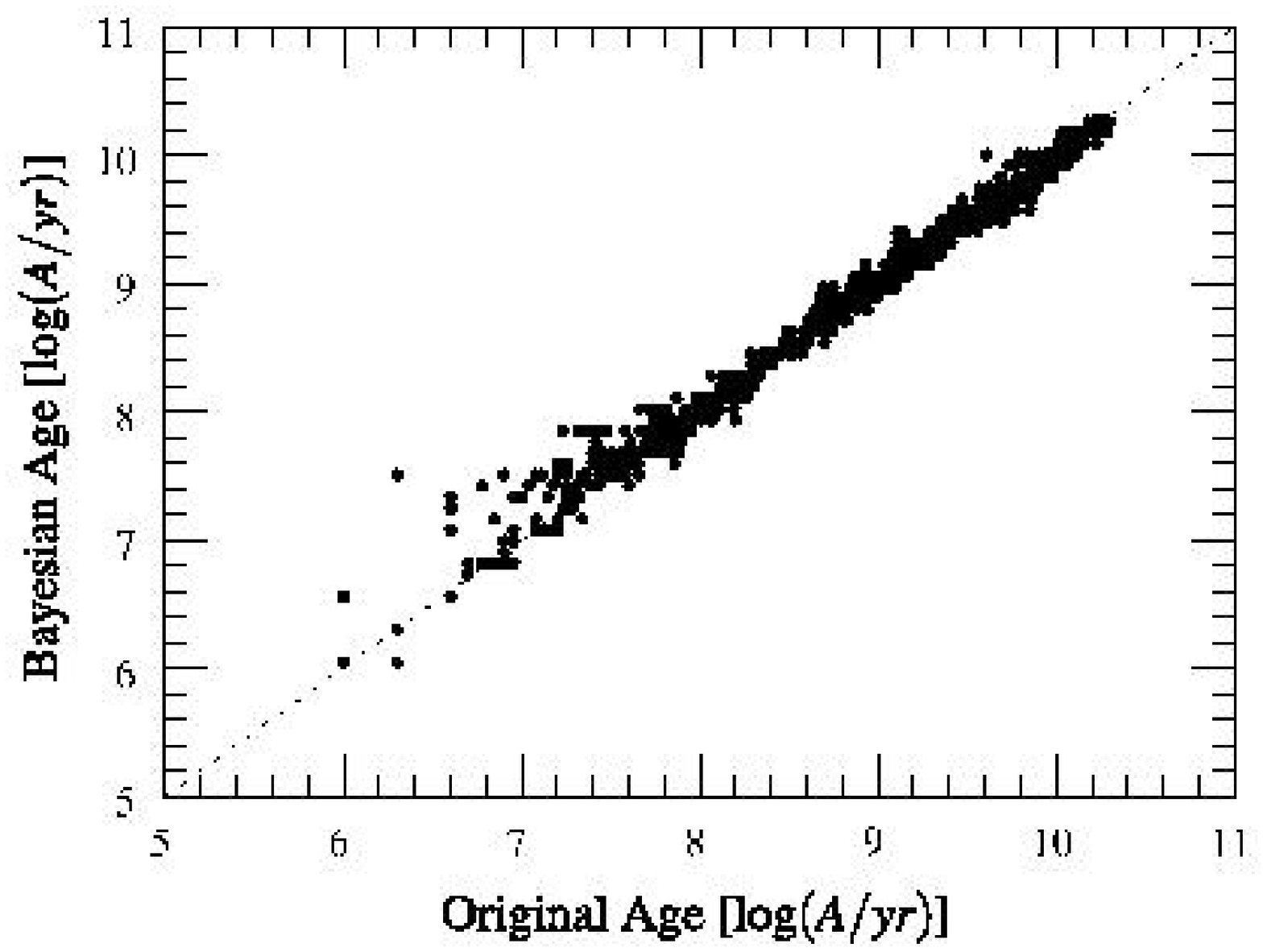}
        \includegraphics[width=8.8cm]{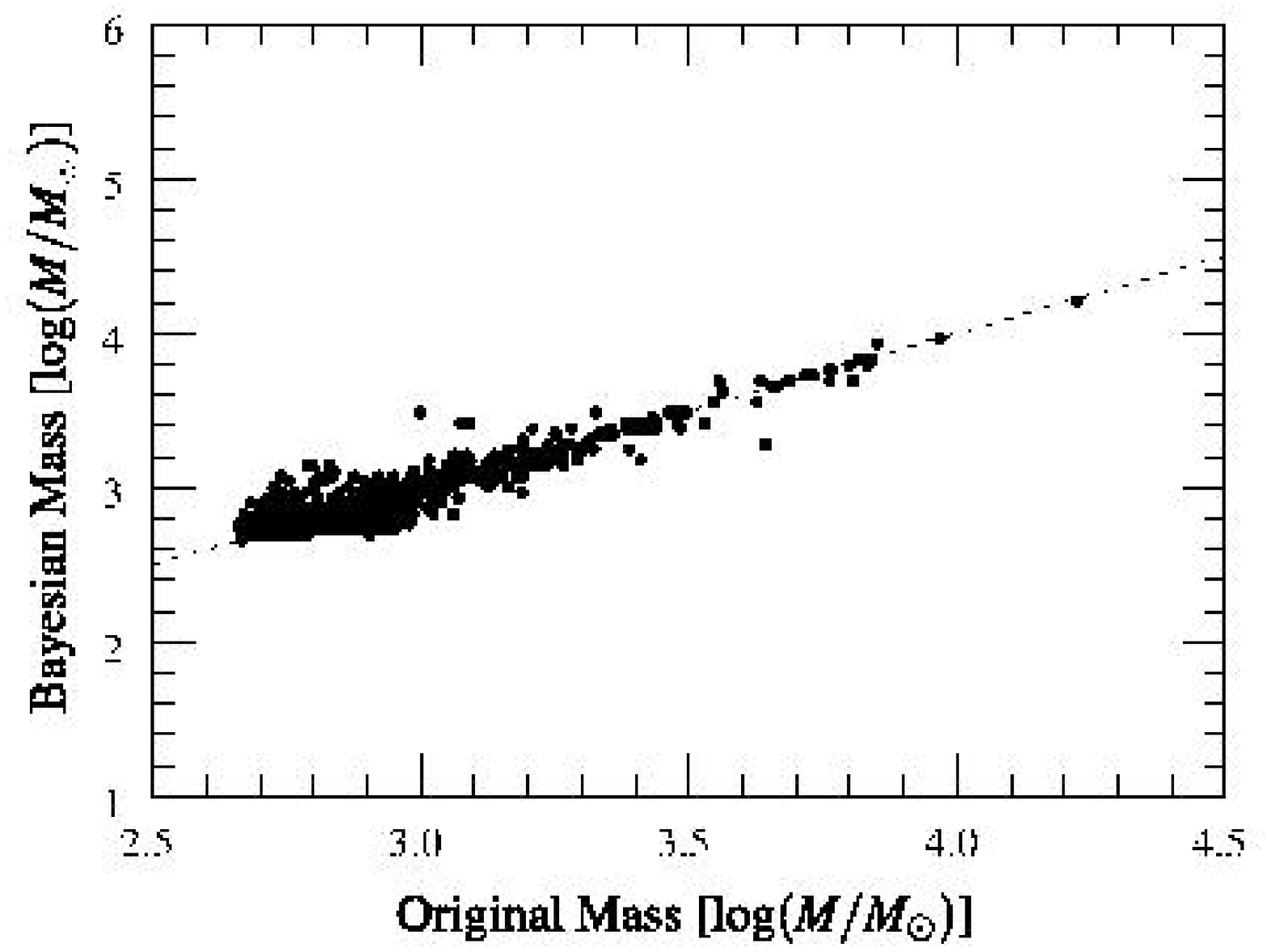}\\
        \includegraphics[width=8.8cm]{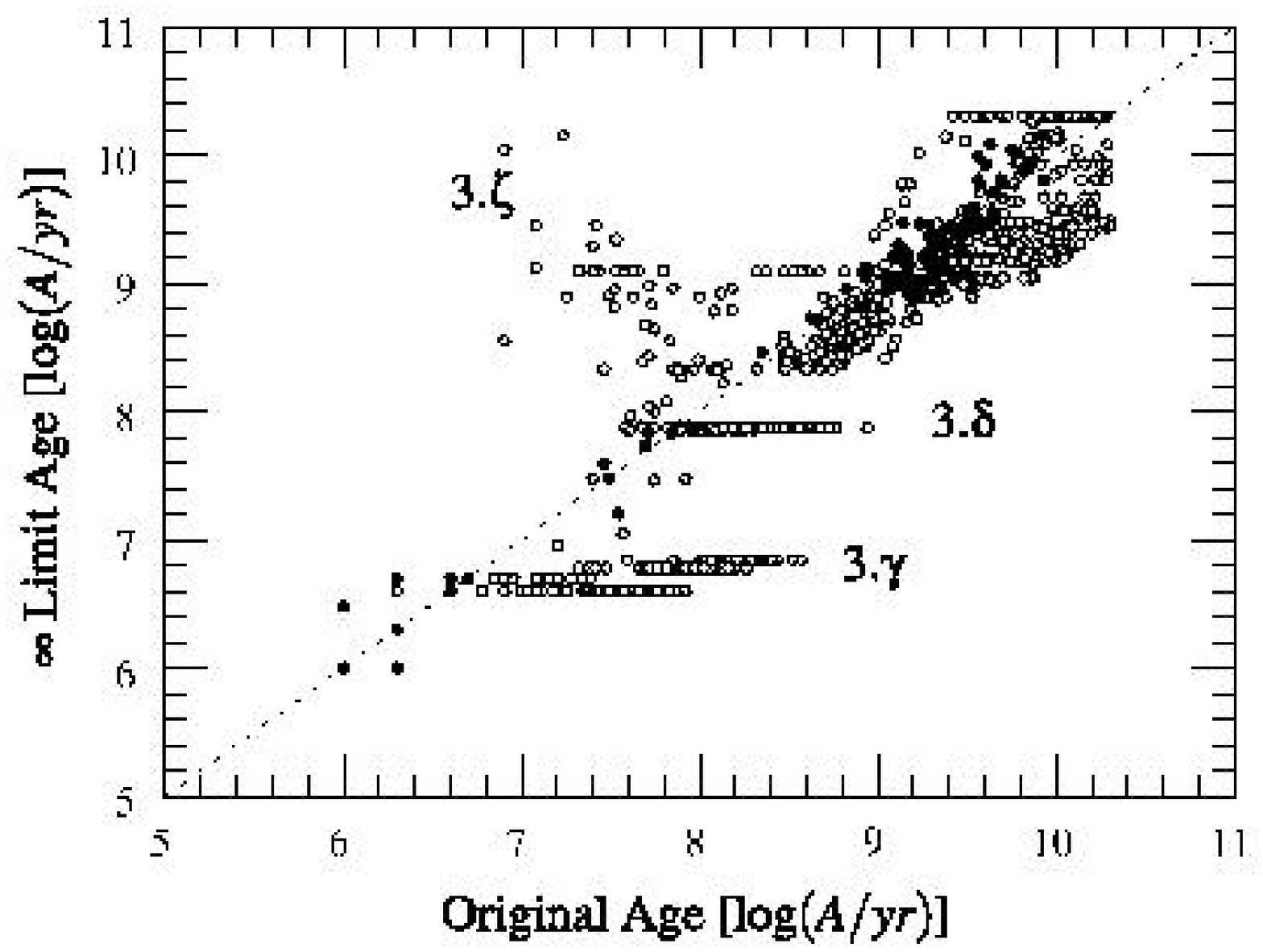}
        \includegraphics[width=8.8cm]{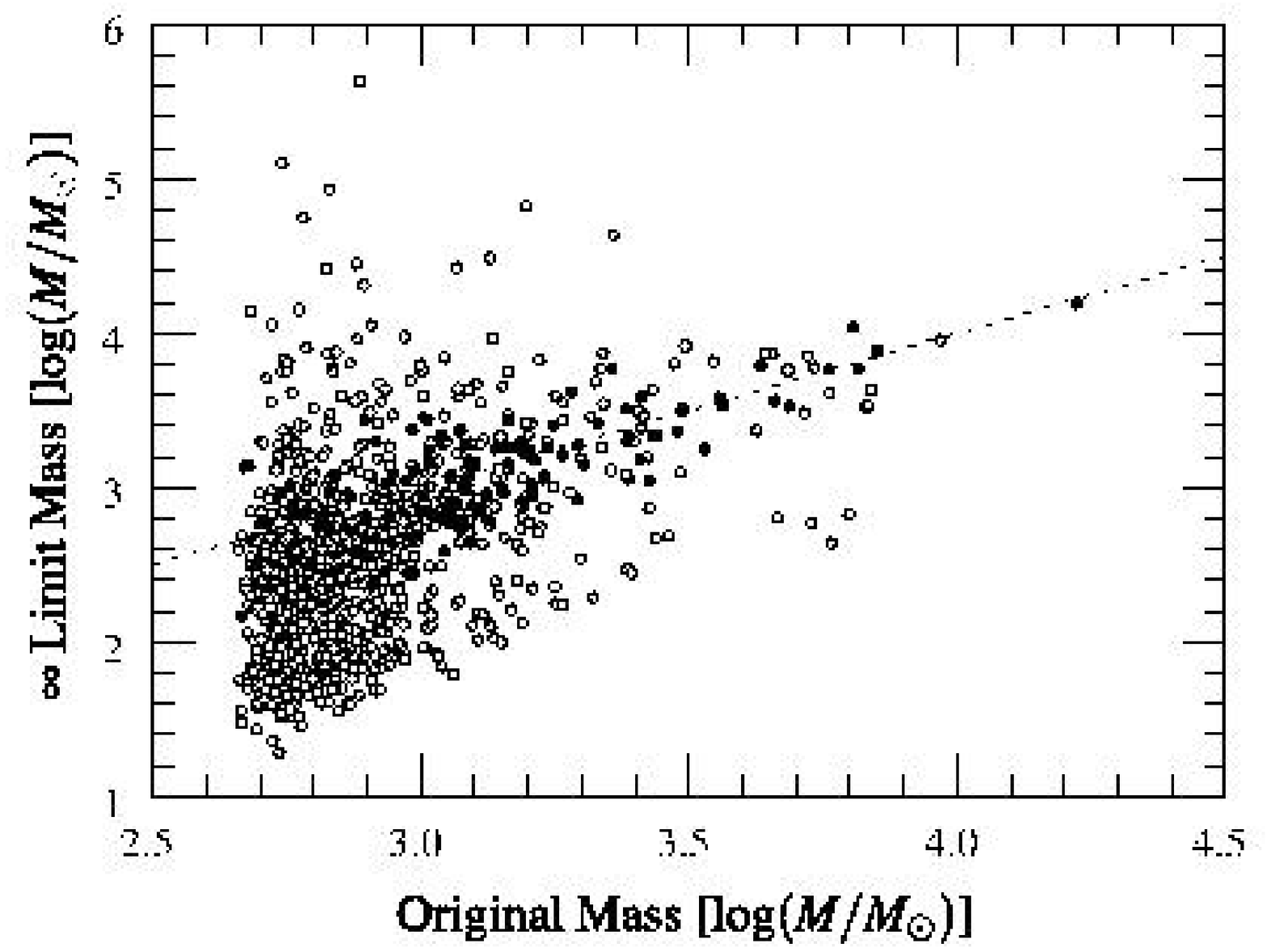}
	\caption{ 
	{
	Estimations based on U, B, V, I and K band fluxes without noise, when
	assuming no extinction (A$_V=0$, ``true'' value).  Estimations from the
	Bayesian approach are given in the two upper panels and standard
	estimation considering an infinite number of stars are given in the two
	lower panels. The x-axes represent, respectively, the original age and
	mass of the sampled populations whereas the y-axes represent the
	estimated values from the two methods.  Dashed lines highlight the
	identity function. Labels are referenced features within the text.  Open
	circles highlight poor fits where the reduced $\chi^2$-values are larger
	than $3\sigma$ confidence range, standard deviation of the $\chi^2$-law.
	}
	\label{fig:UBVIK_nonoise_noAv}
        }
\end{figure*}

{The bottom panels of Fig.~\ref{fig:UBVIK_nonoise_noAv} reveal the inadequacy of 
continuous population synthesis models for the analysis of the
colours of realistic clusters of small masses. }

{The fits to the UBVIK data are poor. $93$\% of the $\chi^2$ values are more
than 3$\sigma$ above the value expected for a good fit if one assumes
observational errors of 0.05 magnitudes ($\sigma$ refers to the standard
deviation of a $\chi^2$-law with 3 degrees of freedom). In other words, no
statistically acceptable match is found in most cases.

Cluster ages are typically underestimated, with a few exceptions.  Errors of $1$
or even $2$~dex are not rare. As a consequence, mass estimates are highly
dispersed. And if one fails to reject clusters with poor fits, the masses are on
average underestimated by $0.3$ to $0.5$~dex.}

\begin{figure*}
        \includegraphics[width=8.8cm]{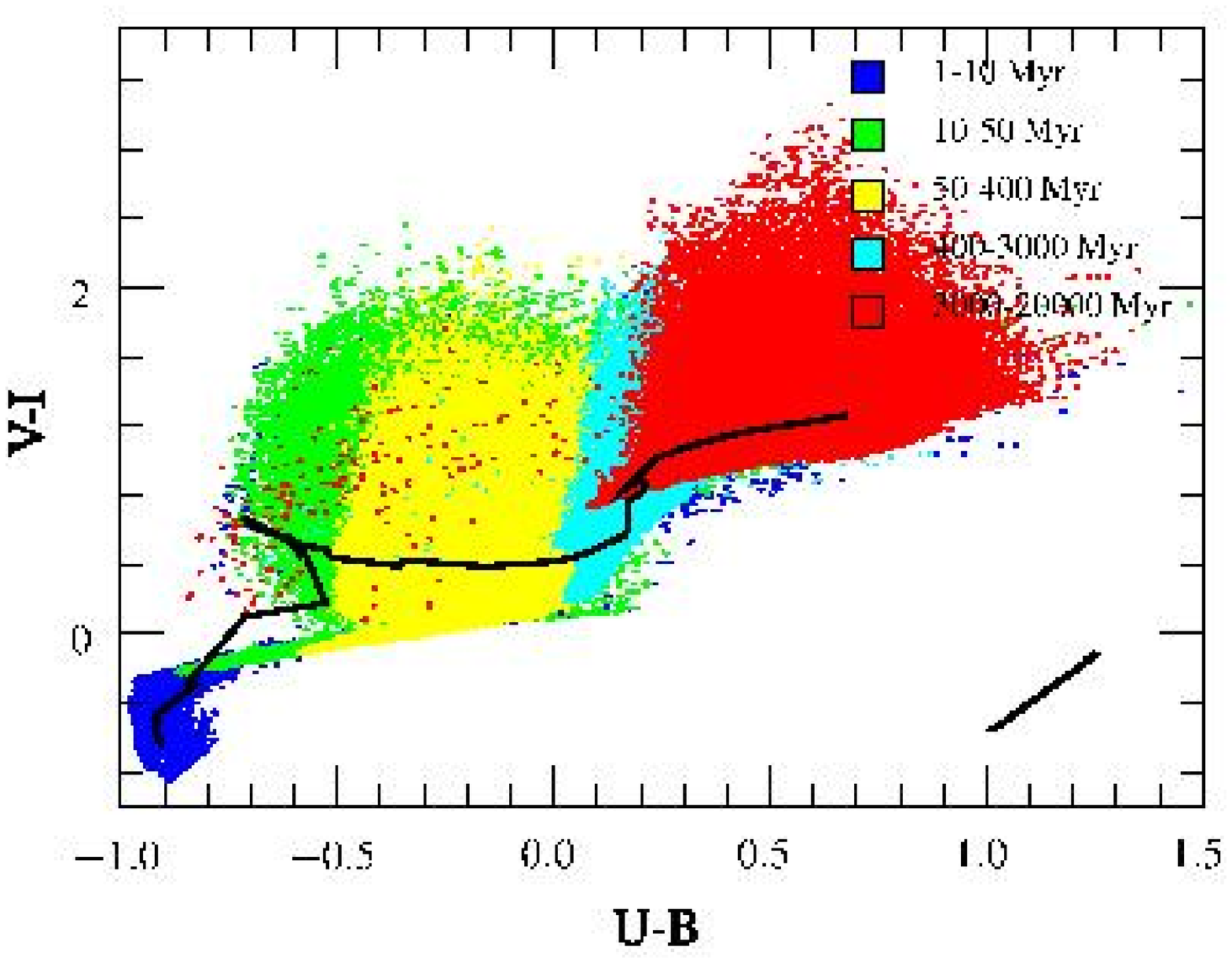}\hfill
        \includegraphics[width=8.8cm]{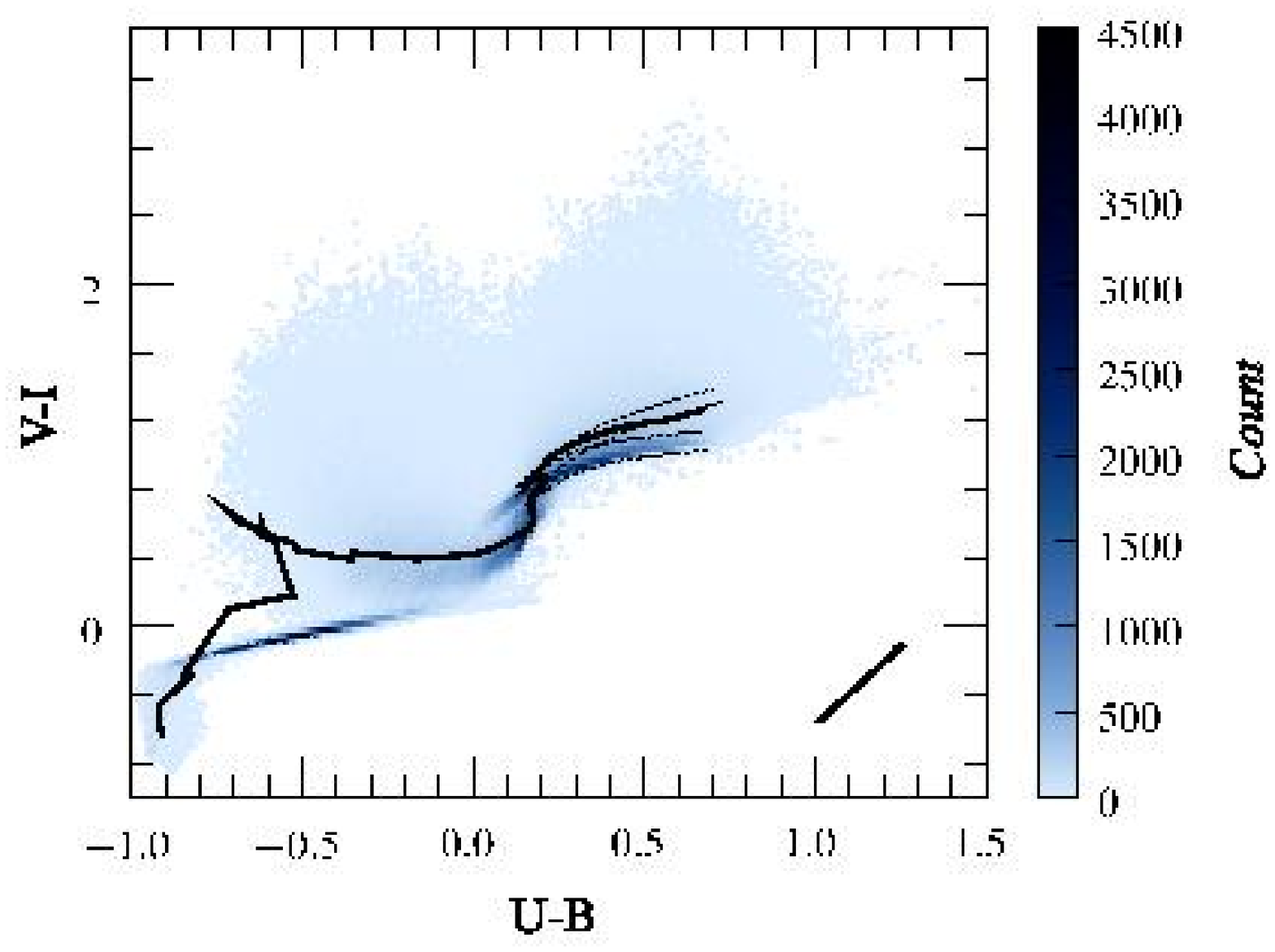}
        \includegraphics[width=8.8cm]{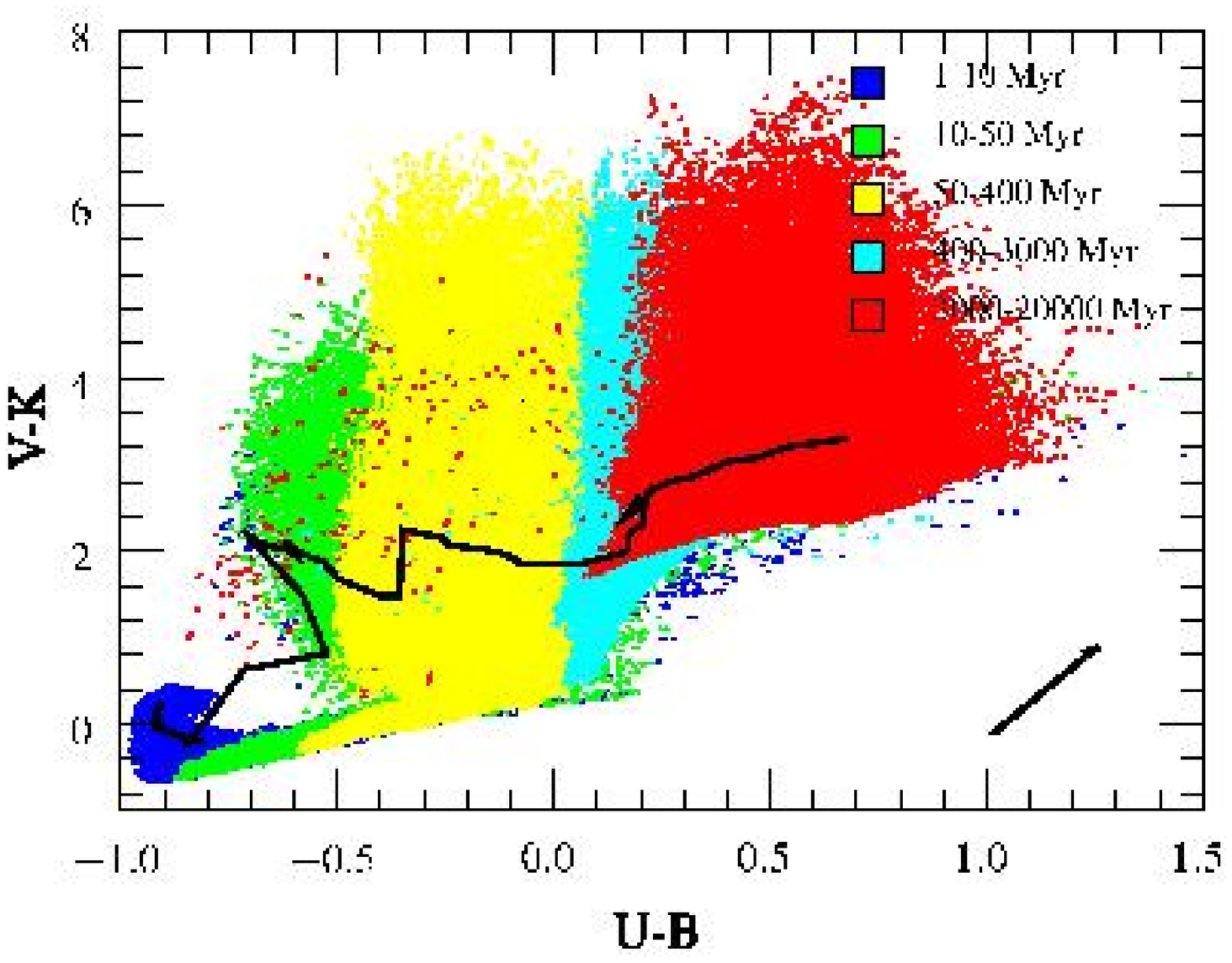}\hfill
        \includegraphics[width=8.8cm]{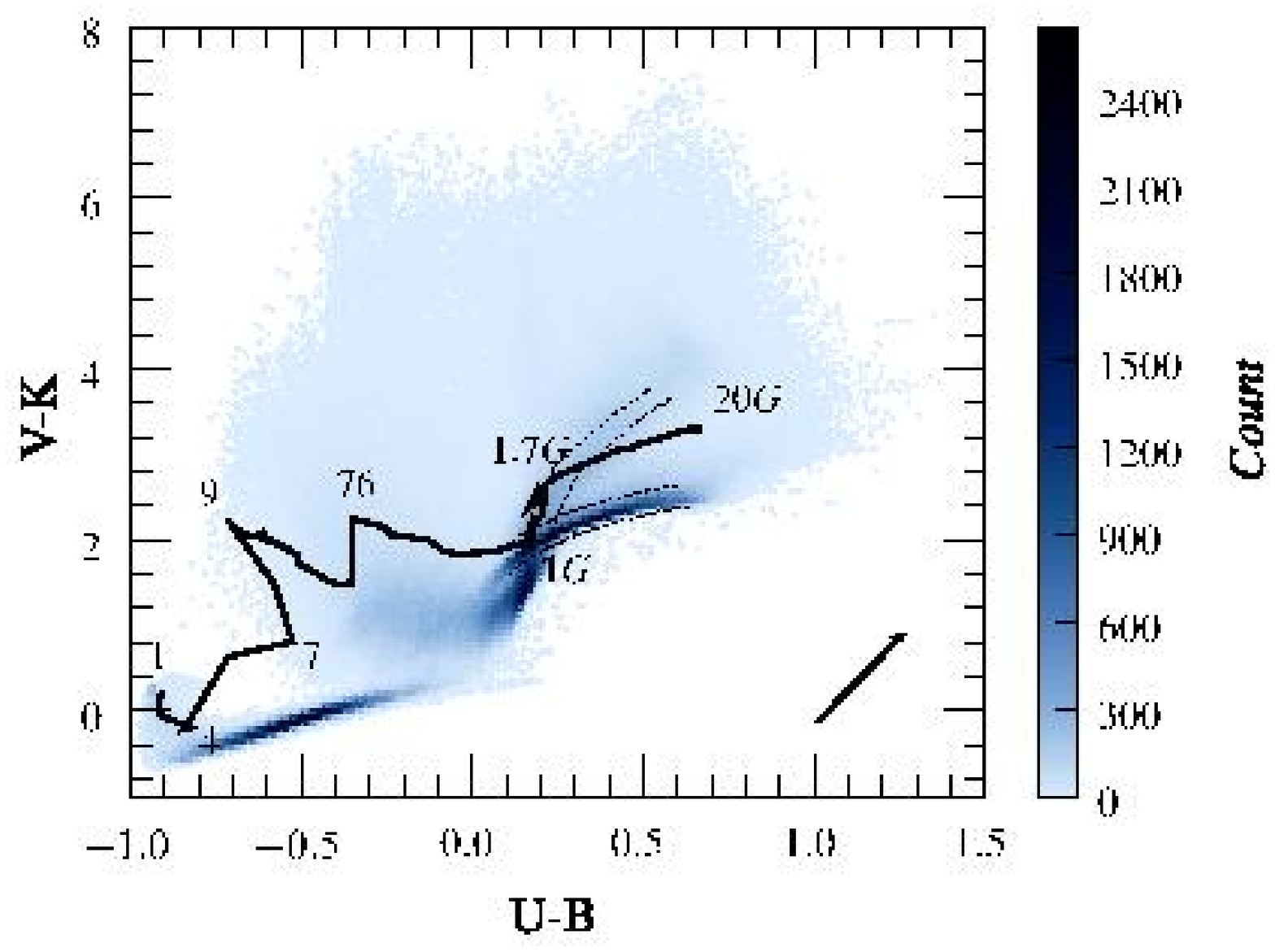} 
	\caption{Photometric properties of stochastic populations compared to
	average predictions at solar metallicity.  On the left hand panels, the
	dots represent coulours of individual clusters contained in our MC
	catalog. Note that there are large regions of overlap between age bins,
	and that older models hide younger models in certain areas of the
	figure. The extinction vector is given by the arrows assuming a Cardelli
	extinction law \citep{Cardelli1989}. On the right panels, the
	corresponding density distributions of the models are shown, assuming
	cluster masses follow a power law with an index -2.  The solid lines
	represent the corresponding age sequences from ``standard'' predictions.
	Numeric labels on the bottom left panel correspond to the age sequence
	from $1$~Myr to $20$~Gyr. {Dotted lines emphasize high density regions
	mentioned in Sect.~\ref{sec:UBVIK_nonoise_noAv}.}
        \label{fig:photoprop}
        }
\end{figure*}

{Fig~\ref{fig:photoprop} helps us understand why derived ages tend to
concentrate around $5-9$~Myr and $80-100$~Myr, in features that we labeled
\ref{fig:UBVIK_nonoise_noAv}$.\delta$ and \ref{fig:UBVIK_nonoise_noAv}$.\gamma$
in Fig.~\ref{fig:UBVIK_nonoise_noAv}, leaving gaps at other ages.  We note that
this remarkable effect is seen in many cluster mass distributions presented in
the literature \citep[e.g.][]{Gieles2009} and has received partial explanations
\citep[e.g.][]{Fall2005}.

Since there is no extinction, the optimization will simply search for the
nearest continuous model in the UBVIK-space.  For intermediate age clusters,
hooks in the locus of continuous models make individual ages more or less
attractive. For instance, the supergiant evolution phase, which occurs around
$9$~Myr in our models, is very attractive and translates into the feature
\ref{fig:UBVIK_nonoise_noAv}$.\gamma$ in Fig.~\ref{fig:UBVIK_nonoise_noAv}.  The
accumulation around $100$~Myr (feature \ref{fig:UBVIK_nonoise_noAv}$.\delta$)
results from the redward excursion of V$-$K when the upper Asymptotic Giant
Branch (AGB) first becomes populated.  The attractive effect of these particular
points is enforced by the fact that the spread of the models is significantly
wider in V$-$I compared to U$-$B (Fig.\ref{fig:photoprop}): moving across wide
parts of the distribution in U$-$B has a lower cost in terms of $\chi^2$
variations than moving along U$-$B.

Although in most cases the best $\chi^2$ values are not statistically acceptable
(again statistically $93$\% of the input-clusters are not well fitted), the
match nevertheless provides a decent age for relatively old populations. 

Finally, objects containing more luminous stars than the continuous models, tend
to be very red. With no extinction, this leads to overestimates of the
population ages, which produces feature~\ref{fig:UBVIK_nonoise_noAv}$.\zeta$ on
Fig.~\ref{fig:UBVIK_nonoise_noAv}. 

Typically, young continuous models are more luminous, per unit mass, than older
ones. Hence any trend in the age estimates will translate into the equivalent
effect on the derived masses. Therefore, this method tends to also underestimate
masses of small clusters.

}

\subsubsection{Bayesian estimates}

{Unlike the standard method, the Bayesian analysis based on stochastic
cluster catalogs returns ages and masses in excellent agreement with the input
values as shown by the top panels of Fig.~\ref{fig:UBVIK_nonoise_noAv}.  As
already mentioned, the single best fit model is exactly the model used to
synthesize the photometry for the analysis. Shown here are the most probable
model properties, after a Bayesian analysis that assumes observational errors of
0.05 magnitudes (although no noise was actually added to the input photometry).

The masses assigned with the Bayesian method are distributed symmetrically
around the input values, except for a cut-off at $500$~M$_{\odot}$ due to the
low-mass limit of our cluster catalog. About $20$\% of the analysed clusters are
potentially affected by this limit. 

The standard deviation of the residuals are of $0.1$~dex in age above $20$~Myr
and $0.08$~dex in mass above $10^3$~M$_{\odot}$ (the age and mass lower limits
are set to avoid statistical biases due to the current limitations of our
MC-catalog).  }

\subsection{When extinction is a free parameter}
\label{sec:UBVIK_nonoise_Av}

\begin{figure*}
        \includegraphics[width=8.8cm]{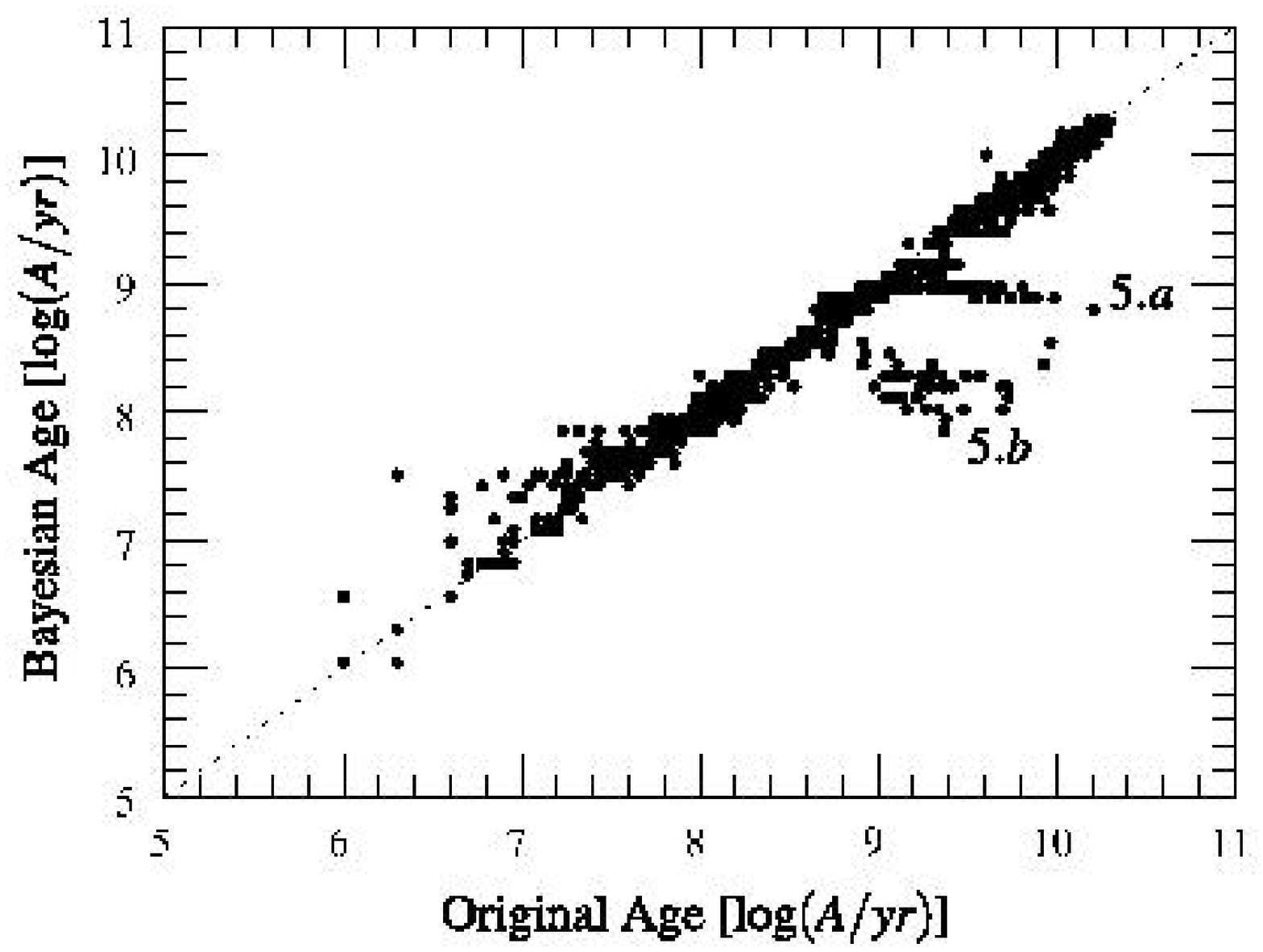}\hfill
        \includegraphics[width=8.8cm]{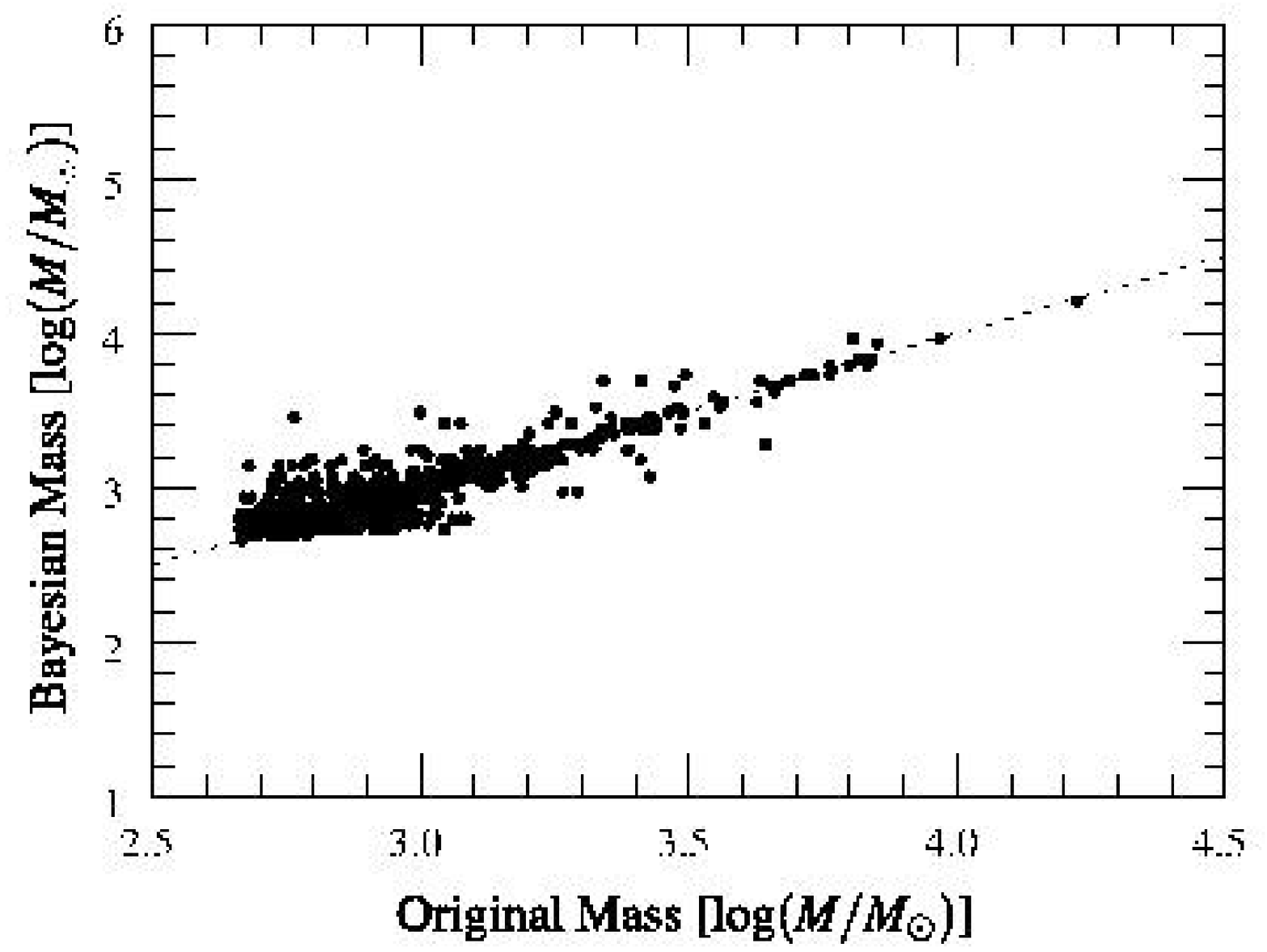}
        \includegraphics[width=8.8cm]{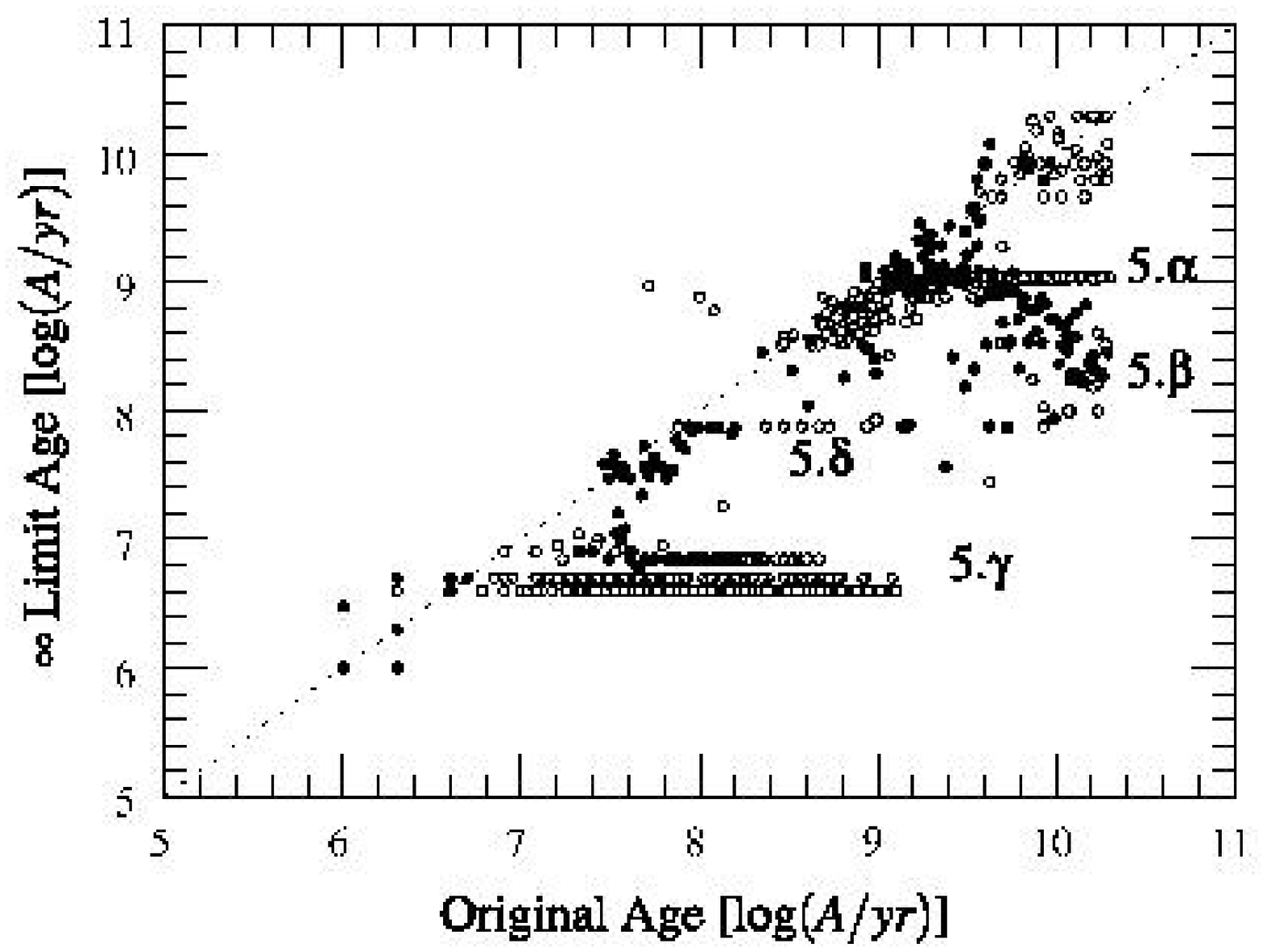}\hfill
        \includegraphics[width=8.8cm]{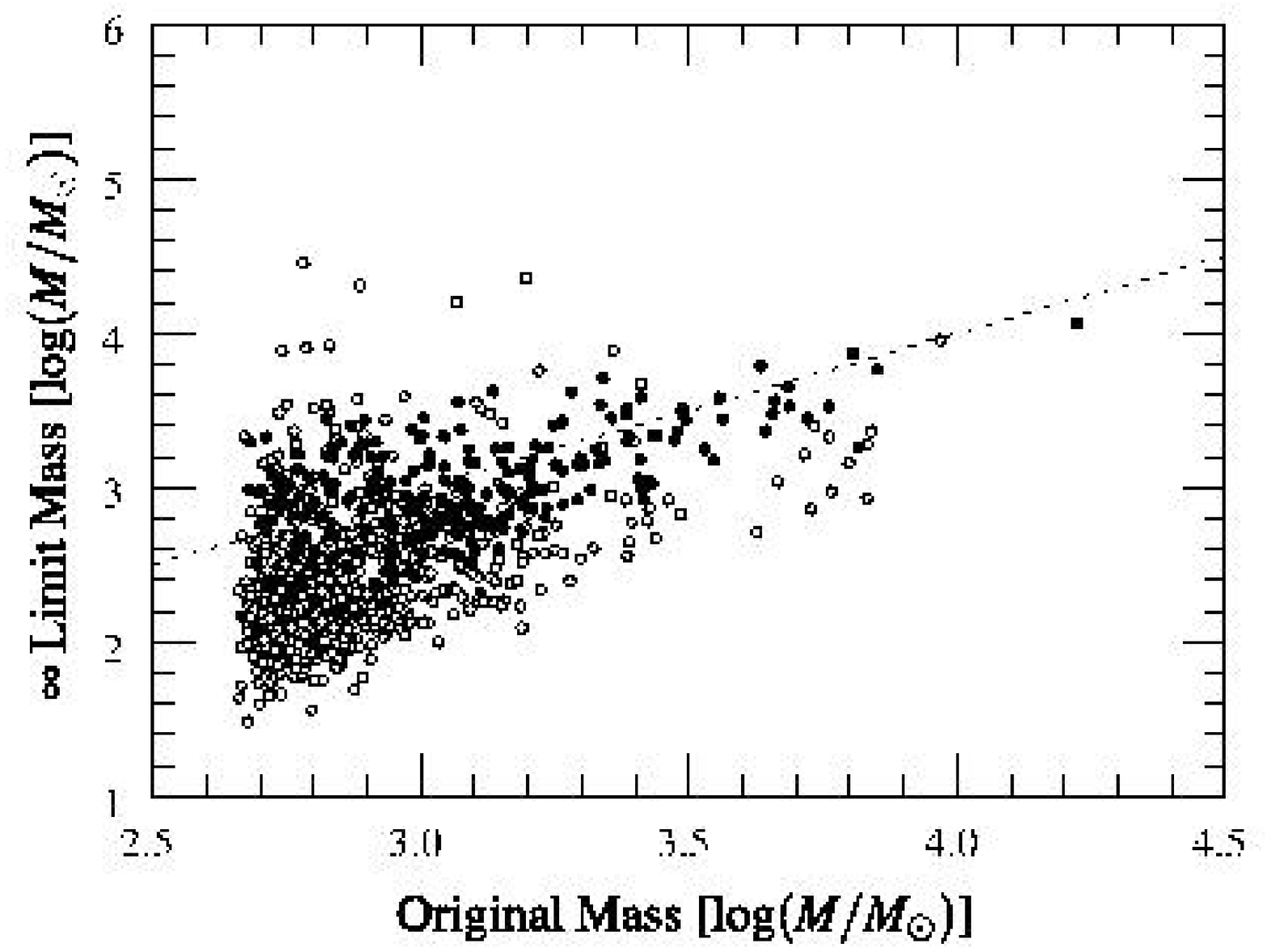}
	\caption{ 
	{Estimates based on U, B, V, I and K band fluxes without noise,
	allowing an extinction A$_V$ from $0$ to $3$.  The conventions and axis
	significations are the same as Fig.~\ref{fig:UBVIK_nonoise_noAv}.  On
	the left-hand side, the age estimations. On the right-hand side, the
	mass estimations.}
	 \label{fig:UBVIK_nonoise}
        }
\end{figure*}

{In the previous section, cluster colours were analysed by comparison with
models during which the amount of reddening imposed to be null in the
optimisation.

We have shown that in most of cases the standard analysis method leads to
under-estimated ages, and masses, even with high quality photometric data across
the spectrum from U to K. The fraction of clusters with underestimated ages was
significantly reduced when adopting stochastic models rather than the classical
continuous ones.

We now assume that no independent information on extinction is available to the
``observer". Hence for the analysis we allow the extinction, $A_V$, to vary
between $0$ and $3$ with steps of $0.2$.  

Figure~\ref{fig:UBVIK_nonoise} shows the results obtained when repeating the
analysis of the reddening-free sample of Sect.~\ref{sec:Input_sample}, this time
authorizing any reddening of the models.  The colours of a reddened sample
are analysed in Sect.~\ref{sec:Av0}.  }

\subsubsection{Estimates from the standard method}

{The bottom panels of Fig.~\ref{fig:UBVIK_nonoise} confirms the inadequacy
of continuous population models for the analysis of realistic clusters of small
masses.  By authorizing the extinction to vary, we do increase the fraction of
clusters whose colours can be fitted satisfactorily with continuous models, but
only to $22$~\%.

Again assigned ages are highly clustered, and most of them are underestimated.
As a consequence, mass estimates are highly dispersed.  

In the standard method, dereddening is the only cost-free way, in terms of
$\chi^2$, to move an observed cluster to the line representing the continuous
models. If we look at the UBVI or UBVK colour-colour planes shown on
Fig.~\ref{fig:photoprop} (left panels), the reddening vector is almost
orthogonal to the loci of stochastic clusters of constant age. Recall that the
input sample is drawn from the stochastic collection even if it is analysed with
the continuous models. Dereddening therefore leads to underestimated ages.

Derived ages tend to concentrate around around $5-9$~Myr, $80-100$~Myr and also
$1$~Gyr, figures that are labeled \ref{fig:UBVIK_nonoise}$.\alpha$,
\ref{fig:UBVIK_nonoise}$.\beta$ and \ref{fig:UBVIK_nonoise}$.\gamma$ in
Fig.~\ref{fig:UBVIK_nonoise}.  Using the direction of the extinction vector in
Fig.~\ref{fig:photoprop} and starting from the bending points of the line of
continuous models, one can define large areas of the colour-colour planes in
which all clusters will be assigned the same ages (with fits of medium or poor
quality). For instance, all clusters with $-0.5<$U$-$B$< 0.3$ and V$-$K$ < 1.5$
(or V$-$I$ < 0.2$) (clusters in the yellow or blue age bins and below the
continuous model line) will be assigned ages of $5 - 7$ Myr, which causes
feature \ref{fig:UBVIK_nonoise}$.\gamma$, very similar to the previously
described \ref{fig:UBVIK_nonoise_noAv}$.\gamma$. Old clusters are preferentially
located on two branches of high model density in our (prior dependent) MC-model
collection.
Those two regions, emphasized by the dotted lines on the
right panels of Fig.~\ref{fig:photoprop}, are crossing each other around
U$-$B$=0$. Clusters in the lower branch, with the higher model density
(at V$-$K$\sim 2$), will be assigned an age of $1$~Gyr and produce feature
\ref{fig:UBVIK_nonoise}$.\alpha$.  The older of these will be assigned higher
extinction values. Clusters near the second and more diagonal branch (V$-$K$>
2.5$) will be assigned ages between $100$~Myr and $1$~Gyr, with a systematic
trend towards younger ages for redder, older clusters (feature
\ref{fig:UBVIK_nonoise}$.\beta$). Note that the accumulation
\ref{fig:UBVIK_nonoise_noAv}$.\delta$, described
previously, is still present but
weaker. Feature \ref{fig:UBVIK_nonoise_noAv}$.\zeta$ has disappeared.

Again, young continuous models are typically more luminous, per unit mass, than
older ones. In most cases, this age effect is stronger than the luminosity
variation associated with the dereddening procedure. Therefore, obtained masses
are usually smaller than the actual masses. There are exceptions however, for
clusters associated with large estimated (i.e. overestimated) amounts of extinction.
}

\subsubsection{Bayesian estimates}
{
Despite the additional degree of freedom associated with extinction,
the Bayesian analysis based on stochastic cluster catalogs generally
returns ages and masses in good agreement with the input values (top panels of
Fig.~\ref{fig:UBVIK_nonoise}).

Above $20$~Myr, $90$~\% of the most probable ages lie within $0.3$~dex of the actual
age (r.m.s. dispersion of $0.1$~dex).
For the remaining $10$\% of the clusters in the sample,
the age estimates present attraction points again, and this leads to
underestimates of up to $1.5$~dex.  The favoured ages are located just below
$1$~Gyr and around $100$~Myr.  The corresponding features are labeled $3.a$ and
$3.b$ in Fig.~\ref{fig:UBVIK_nonoise}. While their locations are not far from
those of $3.\alpha$ and $3.\beta$, their origins are different.

The Bayesian ``attractors" are regions of high concentration in the model
density plots of the right panels of Fig.~\ref{fig:photoprop}.  They do not lie
along the locus of continuous population synthesis models.  Since all models
with magnitudes within about 0.05 of the observations are given large weights in
the computation of the posterior probability distribution, areas of high model
density along the dereddening line are favoured.  An illustration is provided in
Fig.~\ref{fig:ExtinctionEffect} and discussed further in
Sect.~\ref{sec:prospects}.  The clusters affected by this issue are usually
located in regions of low model density of the colour-colour and
colour-magnitude planes; they are assigned positive extinction values.
}

\subsubsection{Reddened input clusters}
\label{sec:Av0}
\begin{figure*}
        \includegraphics[width=8.8cm]{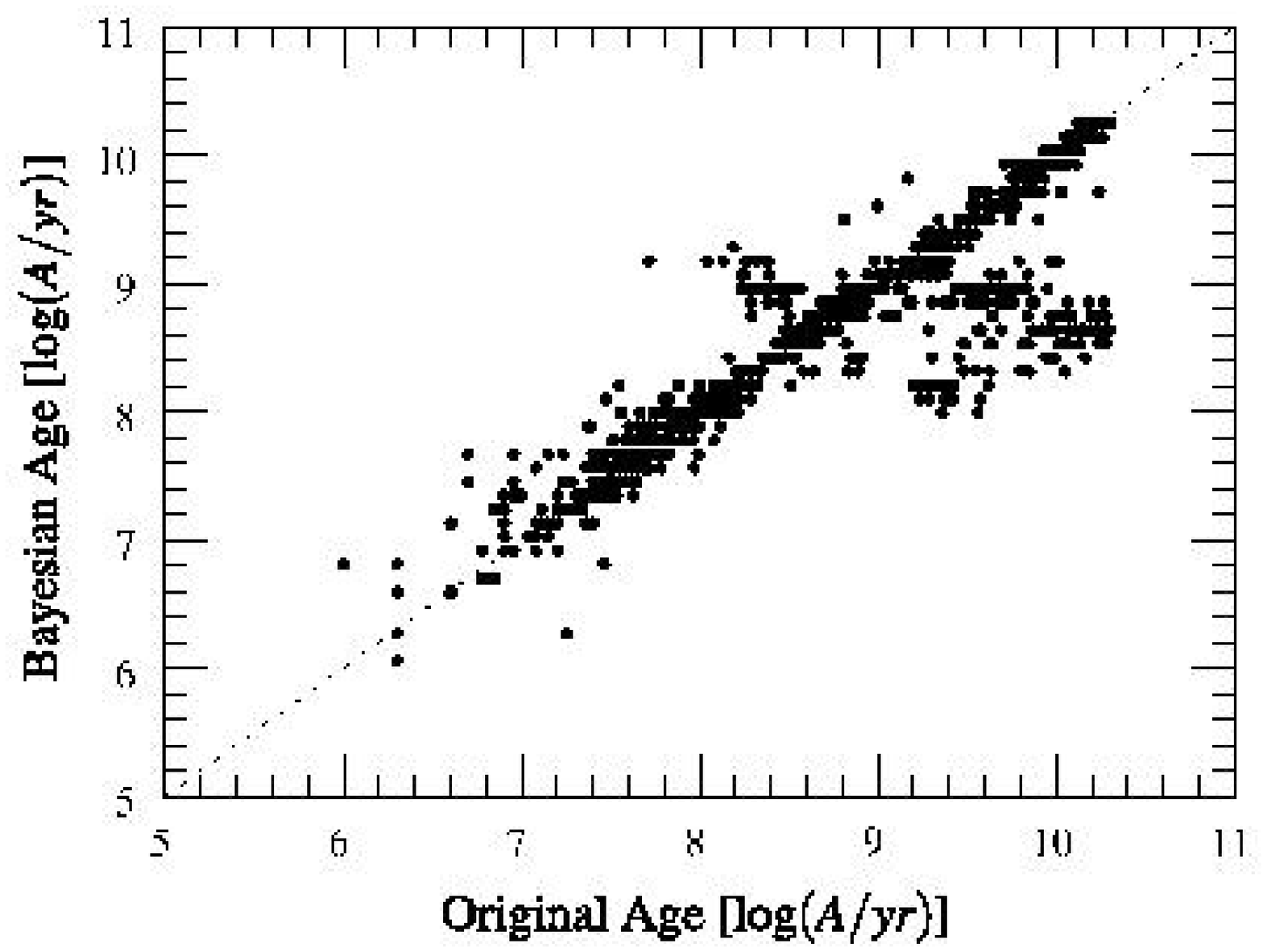}\hfill
        \includegraphics[width=8.8cm]{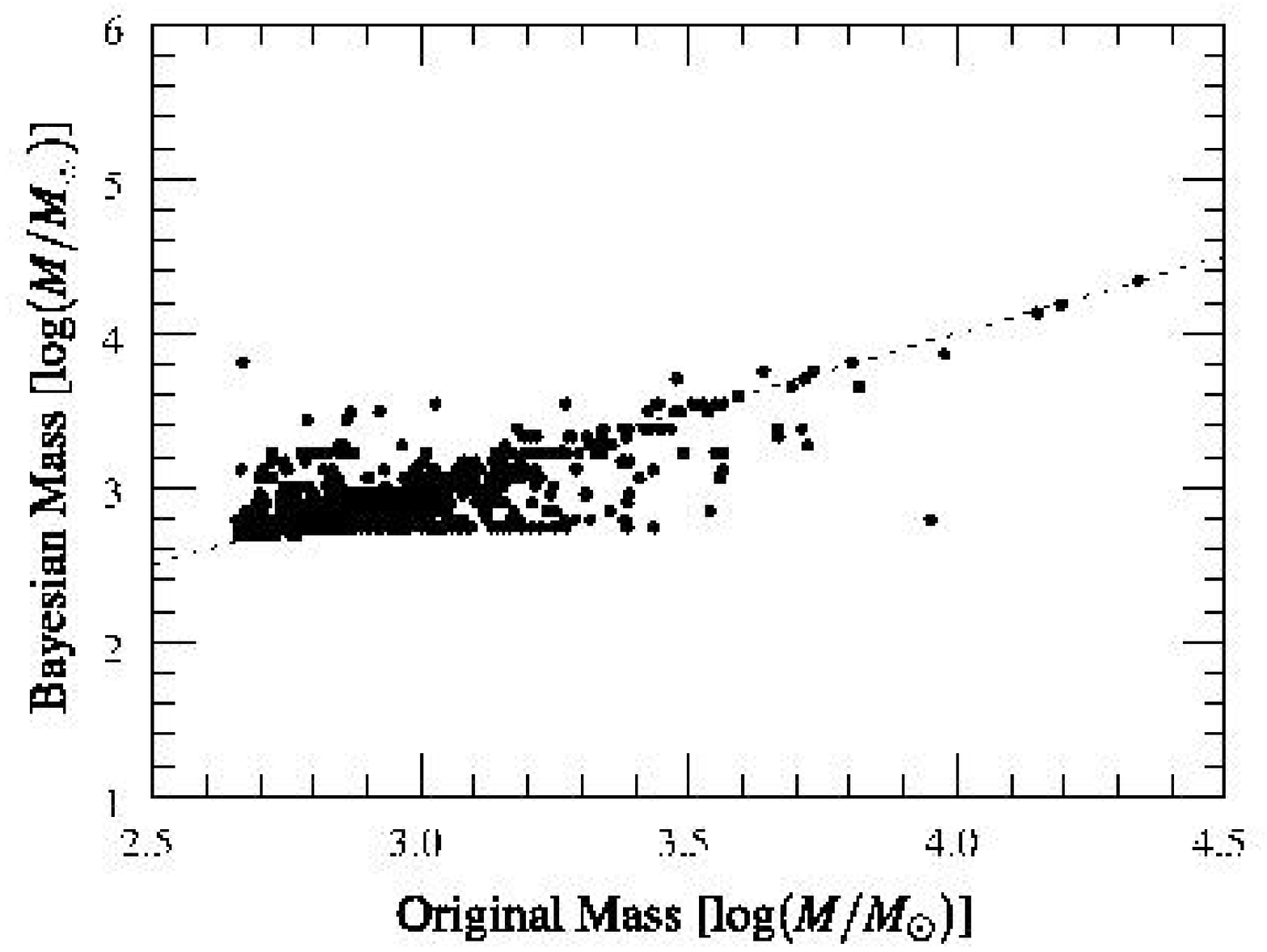}
	\caption{ 
	{Derived age (left) and mass (right) distributions using Bayesian method analysing
	UBVIK data when $1$~mag extinction (assuming Cardelli extinction law) is
	applied to the input population properties.
	This figure compares directly with Fig.~\ref{fig:UBVIK_nonoise}.}
	\label{fig:UBVI_AV0_nonoise}
	}
\end{figure*}

{
As the extinction seems to play a role in the resulting trends of
derived ages and masses from integrated photometry analyses,
we conduct the same experiment as before, where the input sample
photometric dataset was reddened by a factor of $1$ optical magnitude, assuming a
Cardelli extinction law \citep{Cardelli1989}.

Fig.~\ref{fig:UBVI_AV0_nonoise} presents the resulting age and mass
distributions of the derived properties when using the Bayesian method.
$97$\% of the clusters have a correct derived extinction of $1$~mag according to the
Bayesian estimates.
The standard deviations of the residuals are of $0.2$ dex in age above $20$~Myr and $0.1$ dex
in mass above $10^3$~M$_\odot$.
Trend features presented on Fig.~\ref{fig:UBVI_AV0_nonoise} are very similar to
the previously mentioned ones.  A few clusters now have overestimated ages,
because reddening has moved them from the blue side to the red side of a region
of high model density and typically older ages. This region of high model
density becomes a Bayesian attractor for these reddened clusters, while it was
out of bounds in the reddening-free case (reaching this region of high model
density would have required negative extinction corrections).
}

\subsection{Estimates from noisy fluxes}
\label{sec:UBVIK_noise}

\begin{figure*}
	\includegraphics[width=8.8cm]{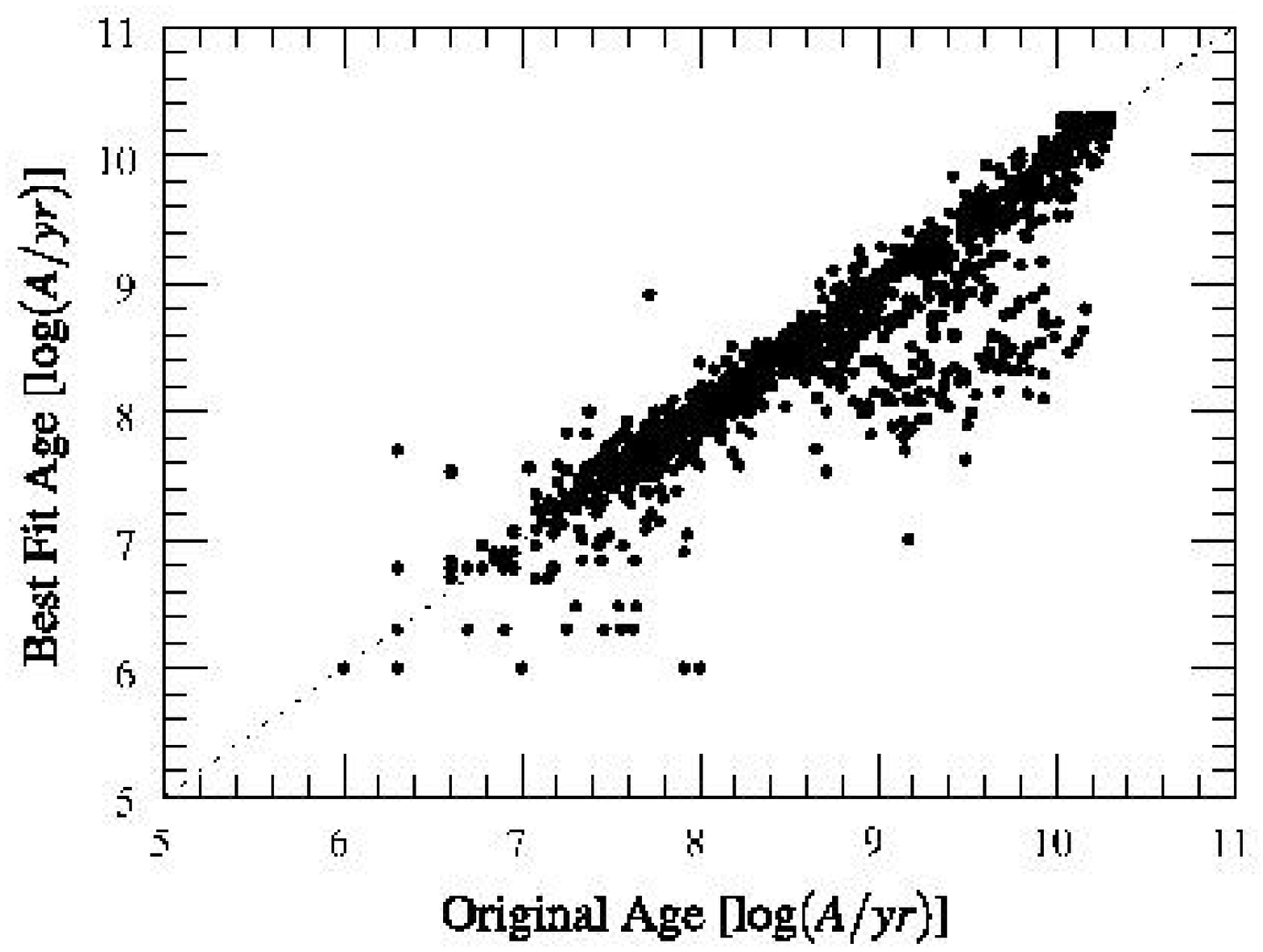}\hfill
	\includegraphics[width=8.8cm]{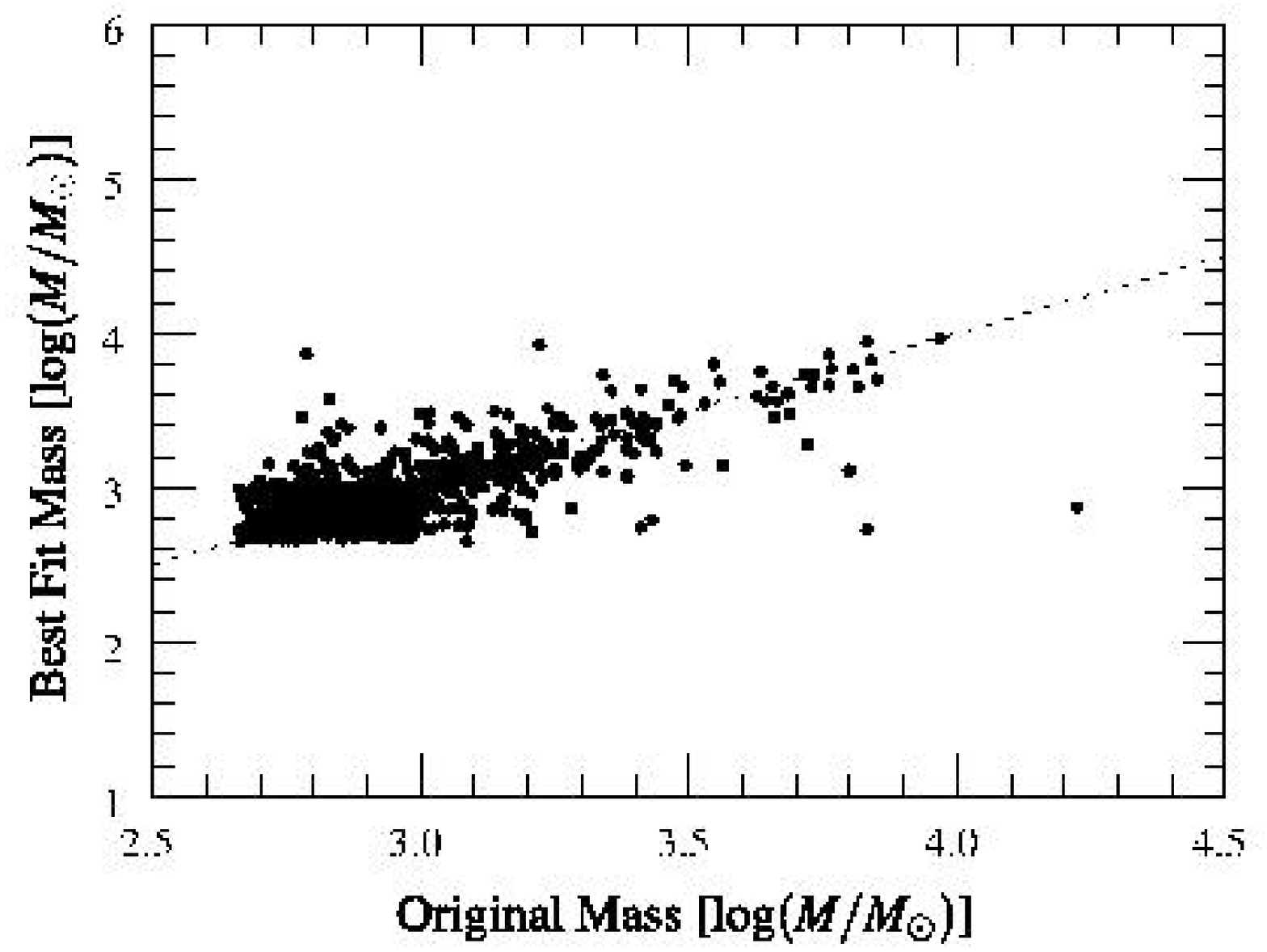}
        \includegraphics[width=8.8cm]{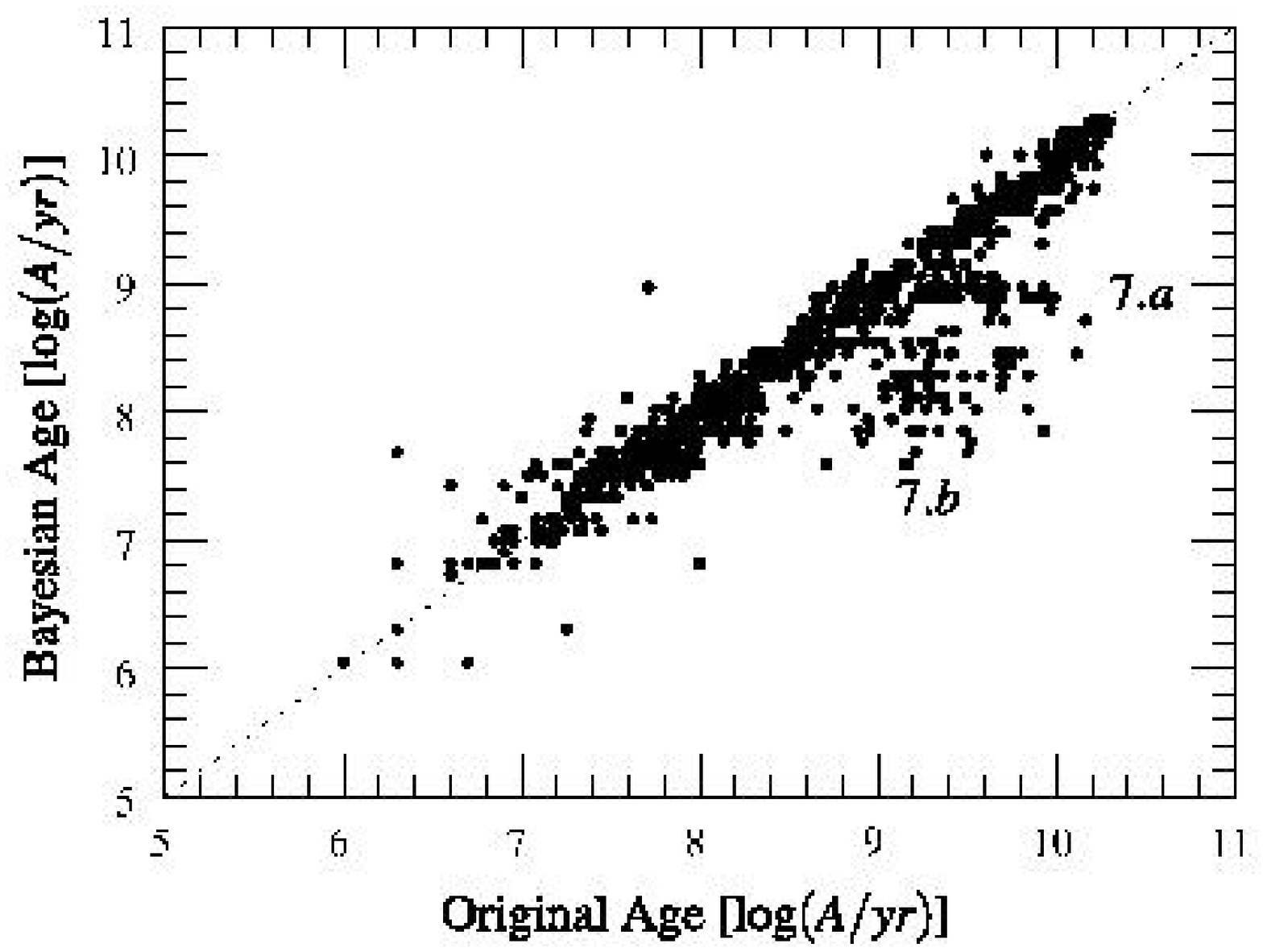}\hfill
        \includegraphics[width=8.8cm]{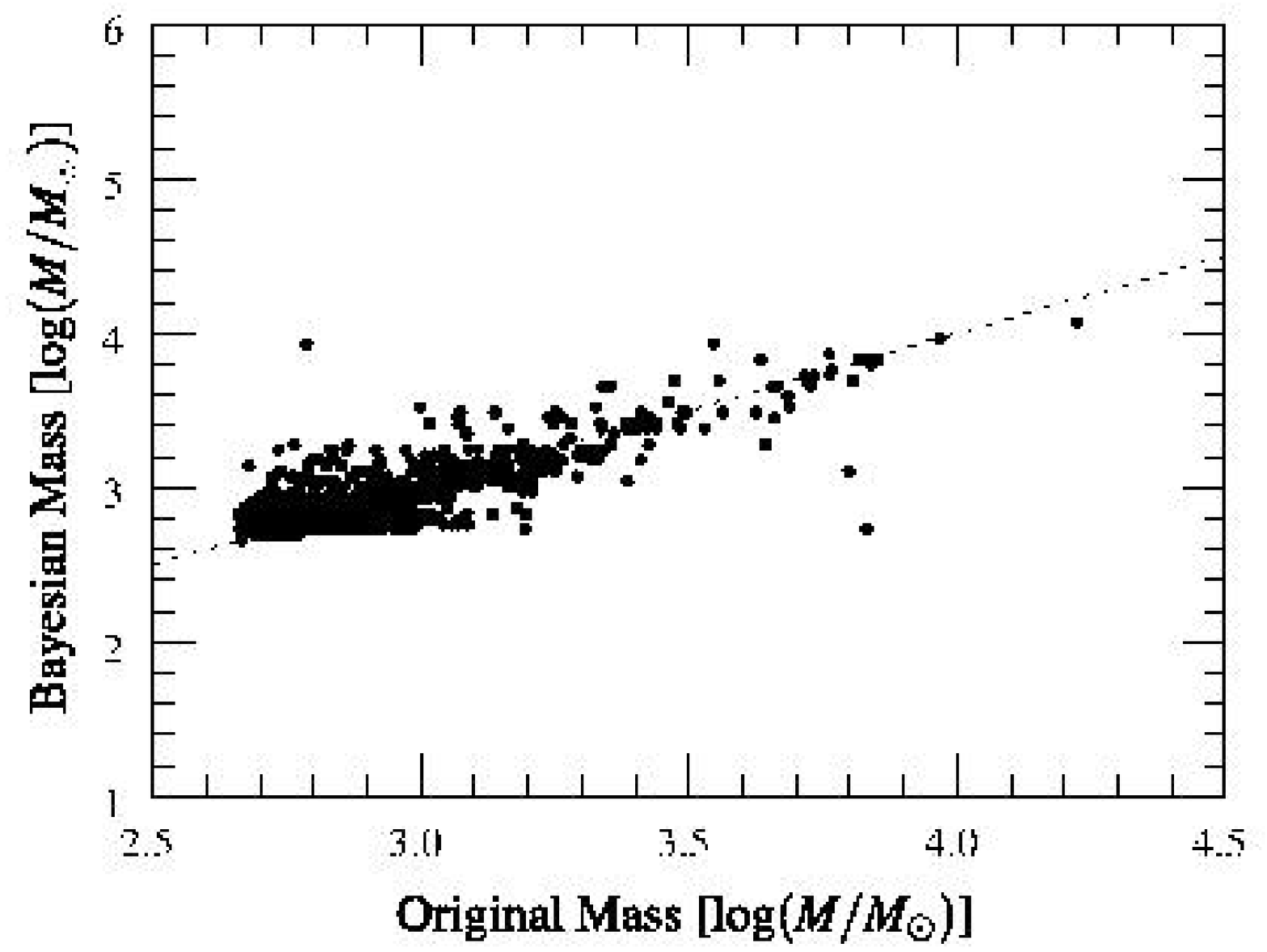}
        \includegraphics[width=8.8cm]{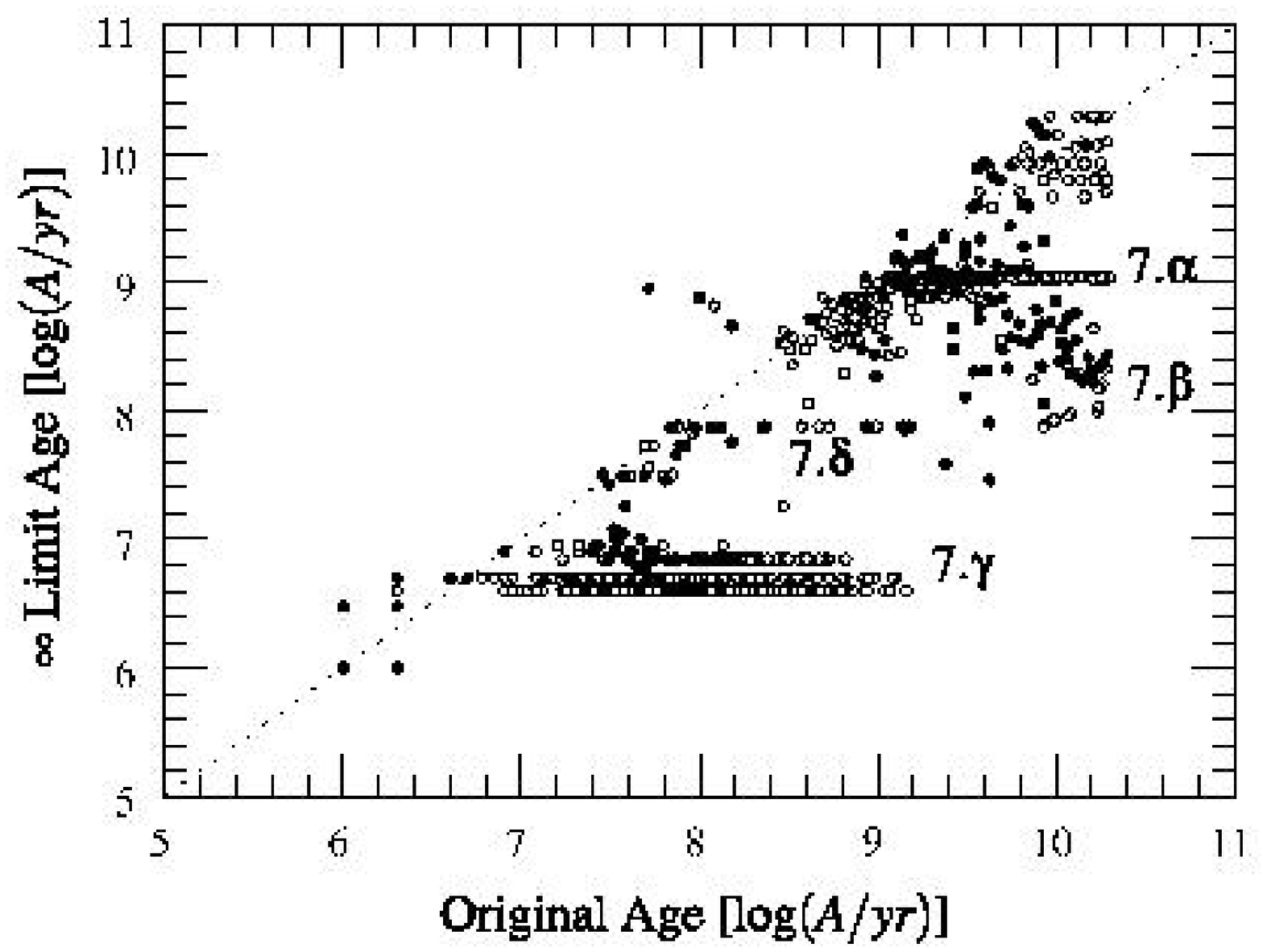}\hfill
        \includegraphics[width=8.8cm]{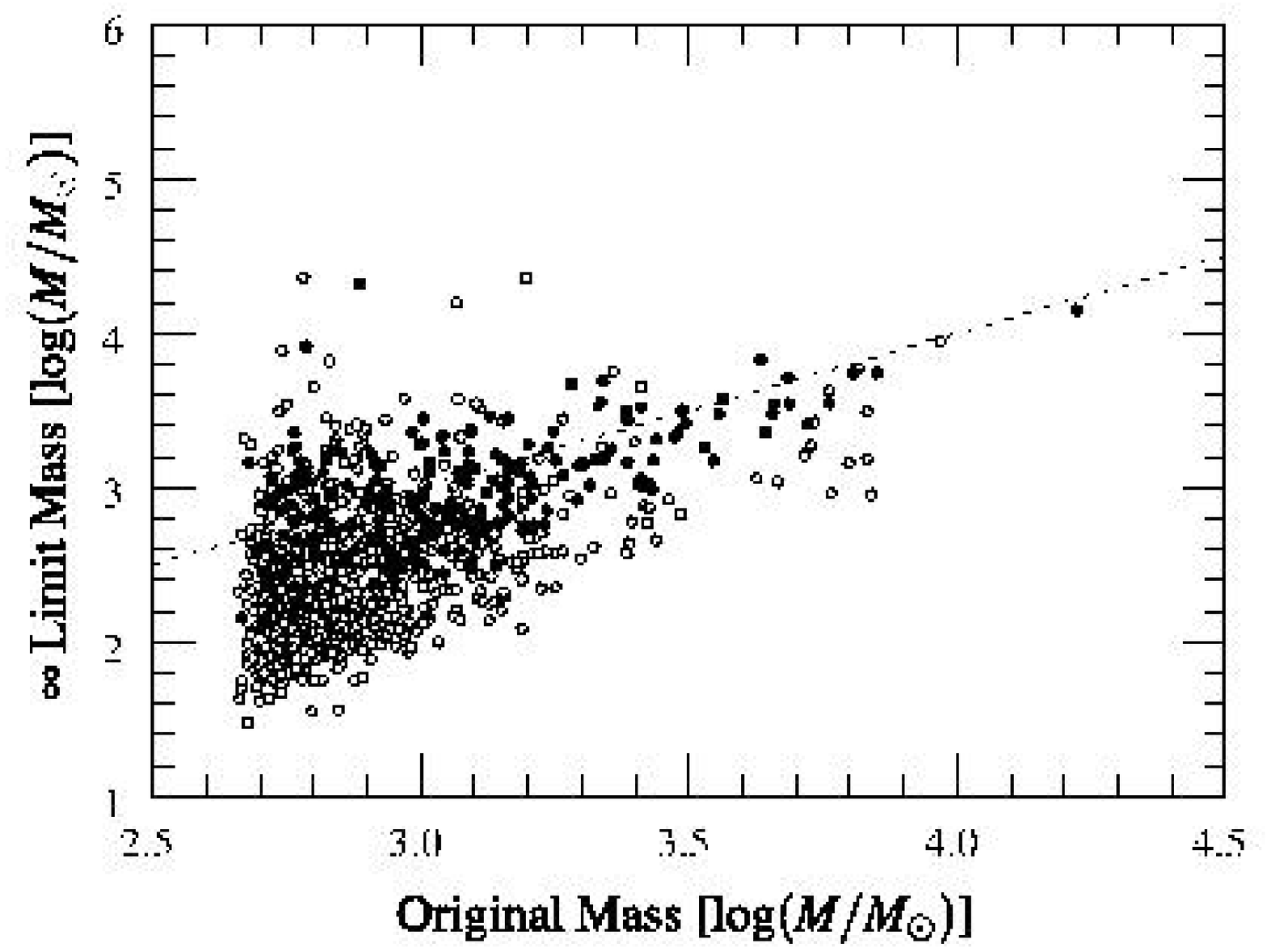}
	\caption{Estimations based on U, B, V, I and K band fluxes with 5\%
	noise added to the photometry inputs, allowing an extinction A$_V$ from
	$0$ to $3$.  From top to bottom, the direct fit of the noised
	populations by our catalog, the Bayesian estimations and the standard
	estimations.  The conventions and axis significations are the same as
	Fig.~\ref{fig:UBVIK_nonoise}.
	\label{fig:UBVIK_noise}
	}
\end{figure*}

We now reproduce the analysis of Sect.~\ref{sec:UBVIK_nonoise_Av} for the same
inputs, but after having added Gaussian noise to the input fluxes ($5$~\%).
Figure~\ref{fig:UBVIK_noise} presents the results. At the top of this figure are
also plotted estimates derived from the single best-$\chi^2$ fit.

Although the inputs are now noisy, the trends obtained remain similar. In
particular, there is no important difference with the noise-free case when the
analysis is based on standard, continuous population synthesis models (bottom
panels): most fits are poor, ages tend to be underestimated, masses are highly
dispersed around the correct values.

Bayesian estimates remain close to the expected values even though the
dispersion has increased. The standard deviation of the residuals are of
$0.15$~dex in age and $0.13$~dex in mass except for models in $6.a$ and $6.b$
features.  Only $13$~\% of the model clusters are assigned underestimated ages.

A pleasant fact is that the single best-$\chi^2$ fit provides results that are
similar to the Bayesian ones {(see discussion in Sec.~\ref{sec:cpu})}.  
Although individual clusters are assigned
different ages, the estimated properties of the sample as a whole are described
similarly with both methods. For 84~\% of the clusters, the single best fit age
and the Bayesian ages are both within $0.3$~dex of the actual ages. Of the 16~\%
that deviate, 3~\% deviate only with the single best fit method, 3~\% only with
the full Bayesian method, and 10~\% deviate in similar ways with both methods.

\section{When there is no K band data}
\label{sec:noKband}

It is common understanding that near-IR light is a better tracer of mass in
galaxies than optical light, and that including near-IR photometry helps break
degeneracies and estimate ages. Particularly illustrative figures on the latter
point can be found in \citet{Bridzius2008}.  However, all these statements rest
on models that are valid only in the limit of large numbers of stars, i.e. not
for most of the real star clusters in the Universe.  This Section re-assesses
the role of photometry beyond wavelengths of $1~\mu$m in the stochastic context
appropriate for small clusters.

\begin{figure*}
        \includegraphics[width=8.8cm]{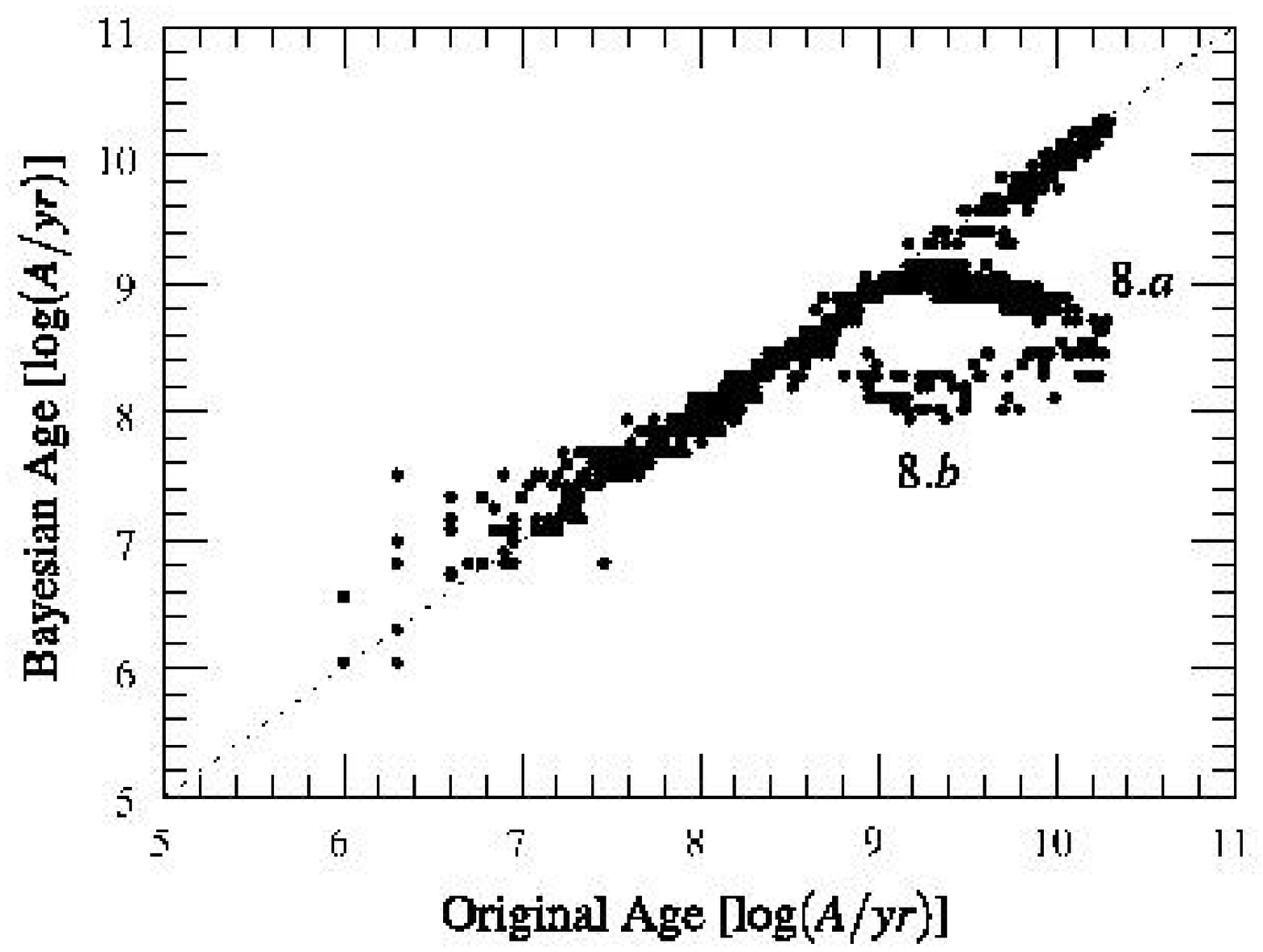}\hfill
        \includegraphics[width=8.8cm]{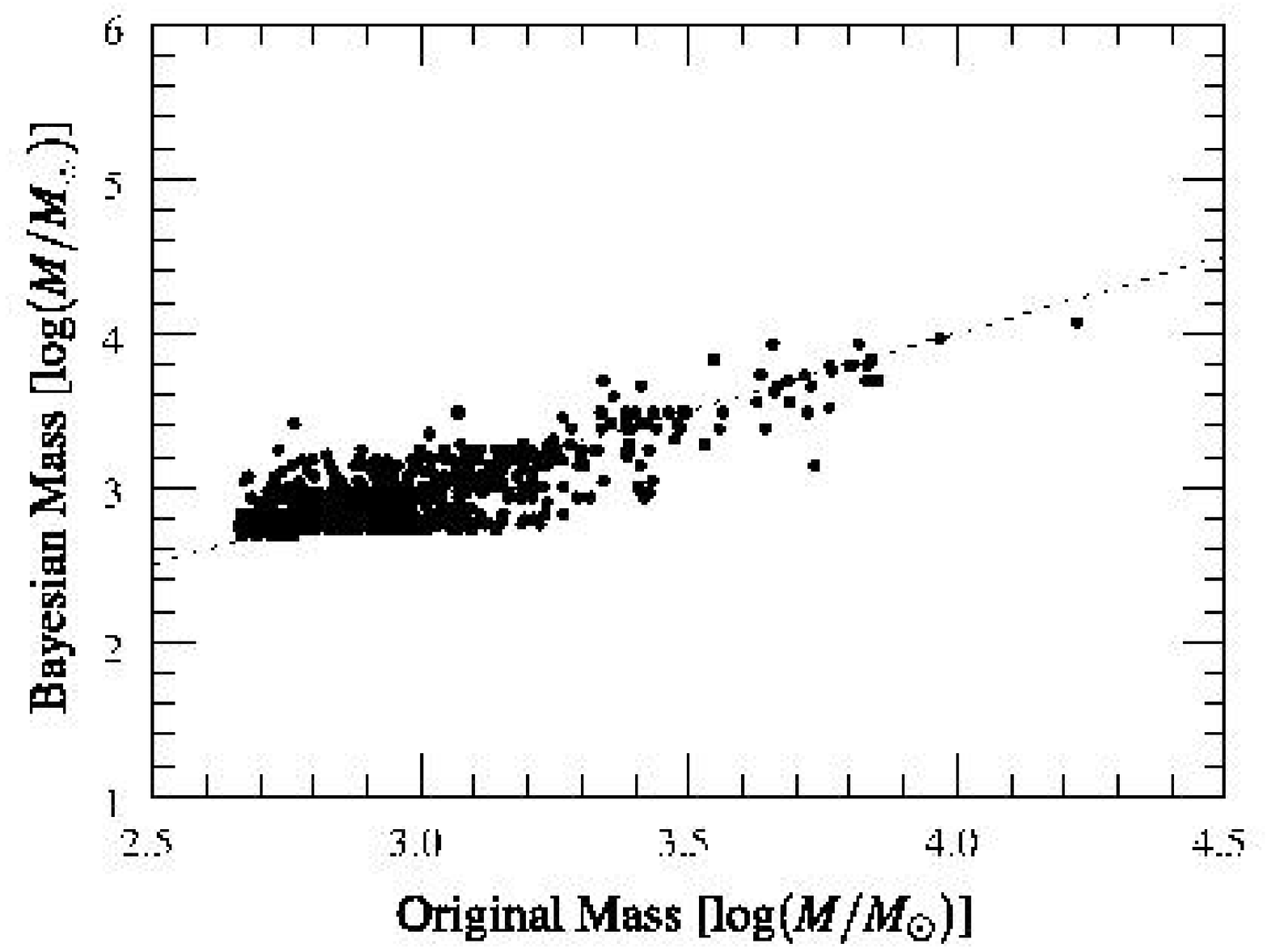}
        \includegraphics[width=8.8cm]{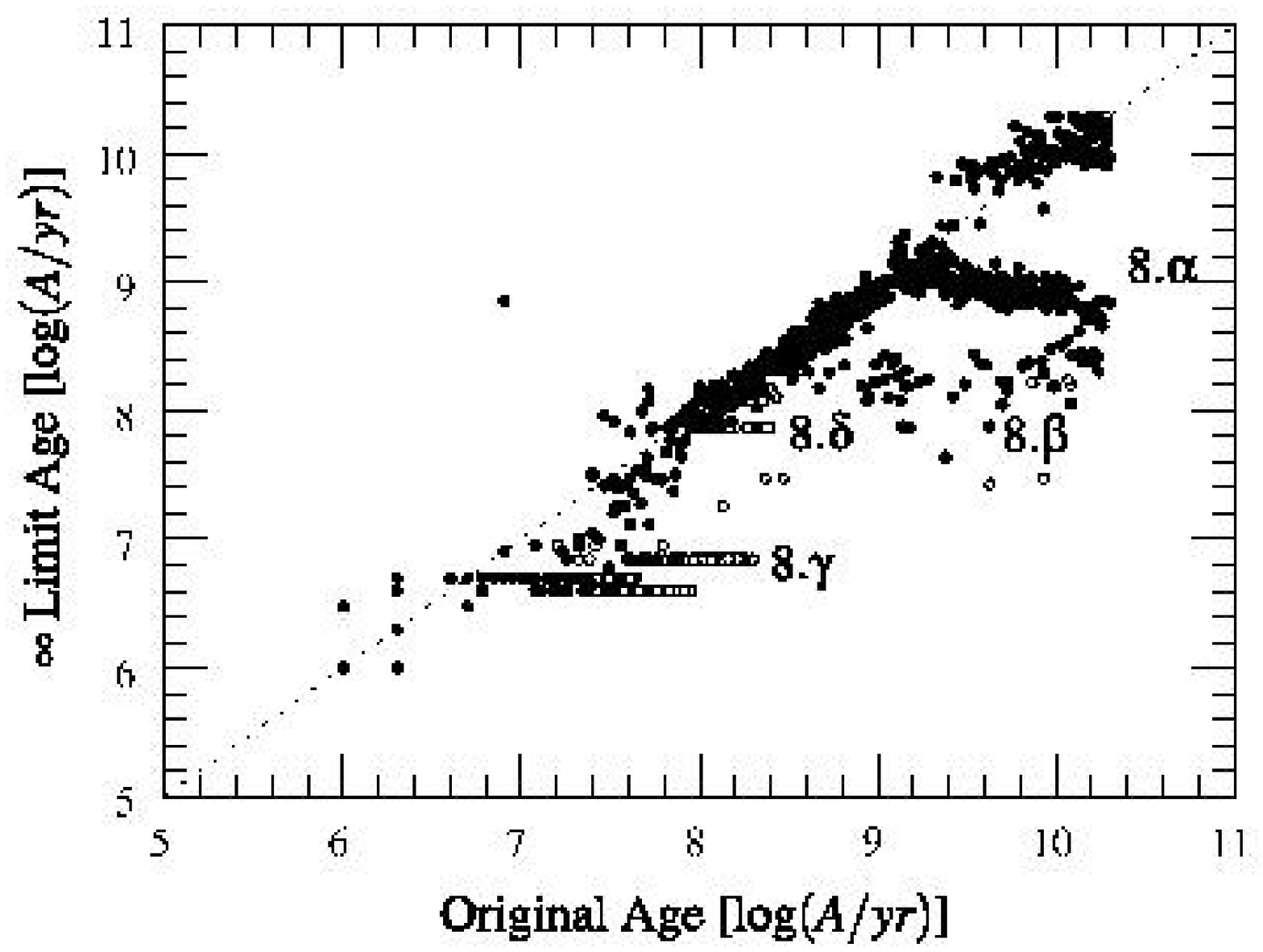}\hfill
        \includegraphics[width=8.8cm]{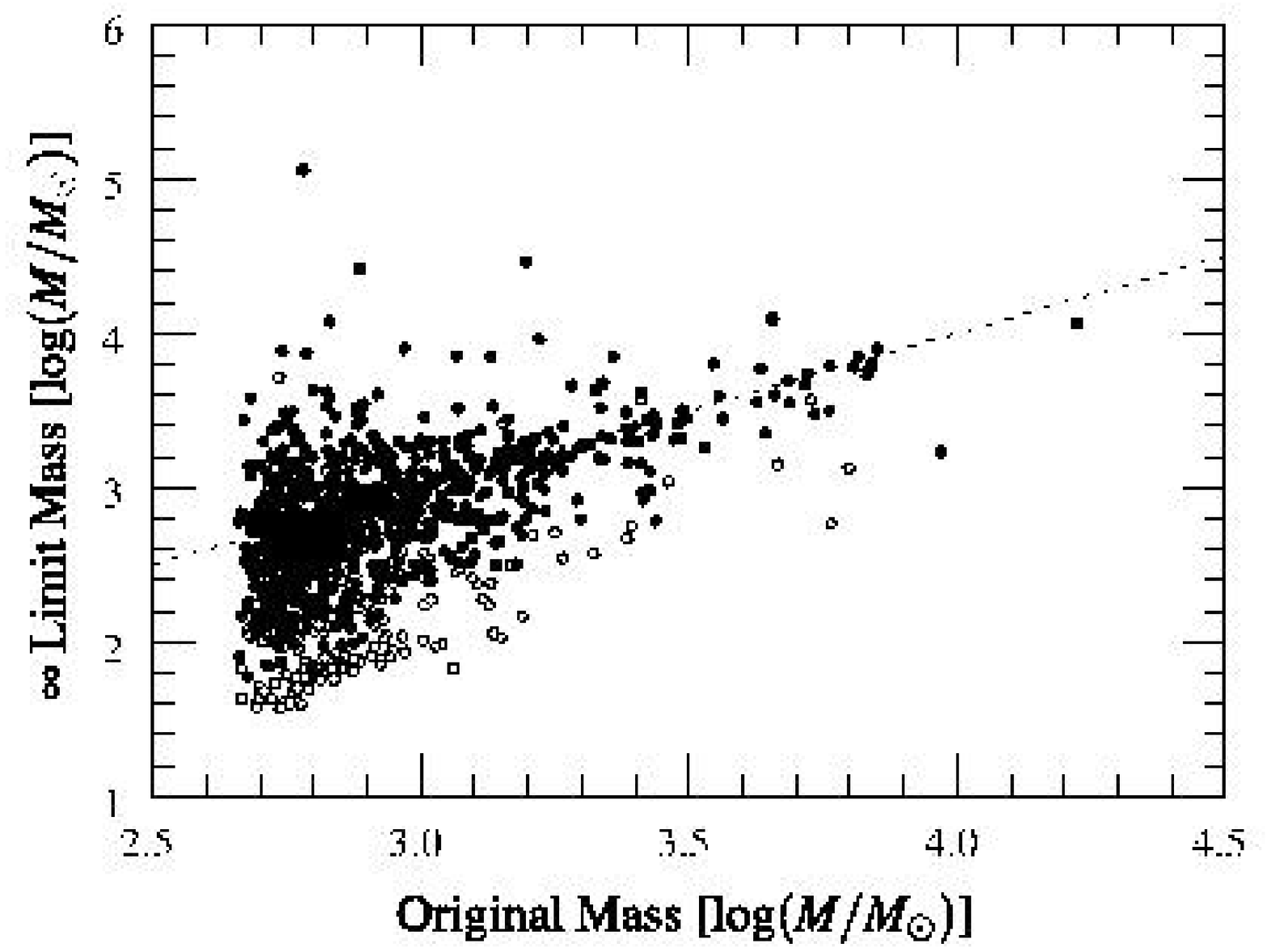}
	\caption{ Same figure as Fig.~\ref{fig:UBVIK_nonoise} but excluding
	information on K band: estimations are based on U, B, V, I bands without
	noise, allowing an extinction A$_V$ from $0$ to $3$.  This Figure
	compares directly to Fig.~\ref{fig:UBVIK_nonoise}.
	\label{fig:UBVI_nonoise}
	}
\end{figure*}

Figure~\ref{fig:UBVI_nonoise} shows the results obtained for the sample of
Sect.~\ref{sec:Input_sample} when using noise-free UBVI input fluxes.
Extinction is a free parameter although, as above, the clusters analysed are not
reddened. Figure~\ref{fig:UBVI_nonoise} compares directly to
Fig.~\ref{fig:UBVIK_nonoise}.

With a smaller number of observational constraints, it is much easier to obtain
statistically acceptable $\chi^2$ values.  $80$~\% of the clusters in the sample
now find an acceptable match among the standard, continuous population synthesis
models (with extinction when convenient). 

A very unusual result is that between about $10^8$ and $10^9$~yr the standard
method (based on continuous models) produces better ages when the K data is
absent than when it is present.  This is due to the relative locations of the
line of continuous models and the regions of high model densities, in colour
space (Fig.~\ref{fig:photoprop}).  At intermediate ages the offset is of one
full magnitude in $(\mathrm{V} - \mathrm{K})$, whereas it remains small in other
colours.  For a cluster that is typically located in the high density region,
the nearest continuous model will produce a decent $\chi^2$ without need for
much reddening in UBVI but large $\chi^2$ values in UBVIK.  The rare but
luminous AGB stars strongly increase the impact of stochasticity on the derived
properties so that one should think about throwing K band data away when the
UBVI-estimated age is between $100$ and $500$~Myr.

At other ages, the standard method applied to UBVI fluxes shows artefacts: a
very strong accumulation is found at estimated ages around $1$~Gyr, and hardly
any cluster is assigned ages between $2$ and $7$~Gyr.  But these artefacts are
not much worse than those seen in the analysis of UBVIK data with continuous
models.

In terms of estimated masses, the distributions around the correct values are
similar whether one uses UBVI or UBVIK data as input to the standard method
(continuous population synthesis).  On average, the masses of the clusters for
which the UBVI fit is satisfactory are underestimated slightly ($0.2$~dex at
$10^3$~M$_{\odot}$).

\medskip

When using the Bayesian analysis based on stochastic models, the loss of the K
band information translates into stronger artefacts in the derived age
distributions. The model-density distribution in colour space plays a more
important role in Eq.~(\ref{eq:proba}) when there is less contrast in the
$\chi^2$ distribution, i.e. when K fluxes are absent. Ages in high density
regions along the de-reddening lines become more attractive. Note that these
artefacts disappear if the amount of reddening is known rather than being a free
parameter.

The artefacts in the age distributions of the standard and the stochastic
method, without the K band, appear oddly similar (left panels of
Fig.~\ref{fig:UBVI_nonoise}).  In fact, the clusters in features $7.b$ and
$7.\beta$ are mostly identical, while the clusters in features $7.a$ and
$7.\alpha$ form two subsets with very little overlap.

One must distinguish three main behaviours for clusters that have ages between
$1$ and $10$~Gyr.  Feature $7.\alpha$ is due to the clusters located in the main
high-density region of model colour-space, i.e. the region that runs parallel to
the line of old continuous models in the lower right panel of
Fig.~\ref{fig:photoprop} but is not superimposed on that line. These clusters
will be assigned correct ages with the Bayesian method, but an age around
$1$~Gyr (and a positive extinction) with the standard method.  Feature $7.a$ is
due to clusters located in the secondary high-density region of colour-space,
which happens to be superimposed on the line of old continuous models (when
there is no K band data).  These clusters will be assigned correct ages with the
standard method, but will be attracted to younger ages with the Bayesian method
(with a positive extinction) because of the contrast in the model-density maps.
Finally, the rarer clusters with larger than typical $V-I$ colours will be
attracted to ages around $100$~Myr (with a positive extinction), both with the
standard and the Bayesian methods (features $7.b$ and $7.\beta$).

As noted earlier, underestimated ages lead to underestimated masses because
stellar populations fade with time and extinction corrections do not quite
compensate for this. In our sample, many of the old clusters affected by the age
artefacts get assigned masses near the lower limit ($500$~M$_\odot$) of our
catalog (see Sect.~\ref{sec:prospects}).

Outside the main age artefacts seen for old clusters (when extinction is a free
parameter), the Bayesian methods recovers ages and masses similarly well whether
or not the K band fluxes are included in the input data.  If we exclude objects
lying in the features $7.a$ and $7.b$ as well as objects affected by the
low-mass limit of our MC catalog, masses recovered with UBVI data sets are not
significantly more dispersed ($\sigma = 0.14$) than those obtained with UBVIK
data sets ($\sigma = 0.13$).  Further aspects of photometric band-pass selection
are discussed in Sect.~\ref{sec:bandselection}.

\section{Discussion}

Several decades have passed since the stochasticity of stellar populations was
first mentioned \citep{Barbaro1977} as a serious issue. With powerful computers,
stochasticity can now be taken into account explicitly when analysing the
properties of unresolved populations.  

Considering stellar populations as stochastic requires some changes in habits,
for instance because mass is not a simple scaling factor anymore: errors on
absolute fluxes (e.g. due to uncertainties on distances) can affect estimated
ages; and observations of only colours can provide some information on the mass.
Clearly, more work is needed to explore the consequences of stochasticity more
exhaustively.

\subsection{Non-stochastic studies : mass and age distributions}

The analysis of the integrated light of small clusters with continuous
population synthesis models produces strong age-dependent artefacts in the
derived ages, and a broad distribution of random errors in the derived masses. 

The good news is that there is no large systematic offset between estimated and
real masses {\em if} the photometric data are of good quality {\em and if}
observations that cannot find a statistically acceptable match among the
continuous models are rejected. When the data have larger errors, fewer model
fits are rejected. Then, a systematic trend appears in addition to the random
errors: masses are underestimated on average. The value of the offset depends on
the photometric pass-bands available and on the existence (or not) of
independent information on extinction. The average error is of $0.3$~dex for a
sample of clusters of masses around $10^3$~M$_{\odot}$ observed in UBVIK, when
extinction is treated as a free parameter.

The above means that mass distributions of large samples of clusters based on an
analysis with continuous population synthesis models are probably not too
strongly biased. Several empirical determinations in the literature
\citep[see][and references therein]{Rafelski2005} favour $M^{-2}$ mass
distributions, and we used such a law as a prior in our Bayesian analysis. We
find no immediate reason to apply a correction to this result, although a
detailed investigation of each individual dataset in the literature would be
worthwhile.  If, for instance, small masses are systematically underestimated by
$0.3$~dex while large masses are not, the power law index of the mass
distribution would require a small correction\footnote{ We have not investigated
any systematics for large cluster masses.  See \citet{Anders2004} for that
regime.}.

\begin{figure*}
        \includegraphics[width=8.8cm]{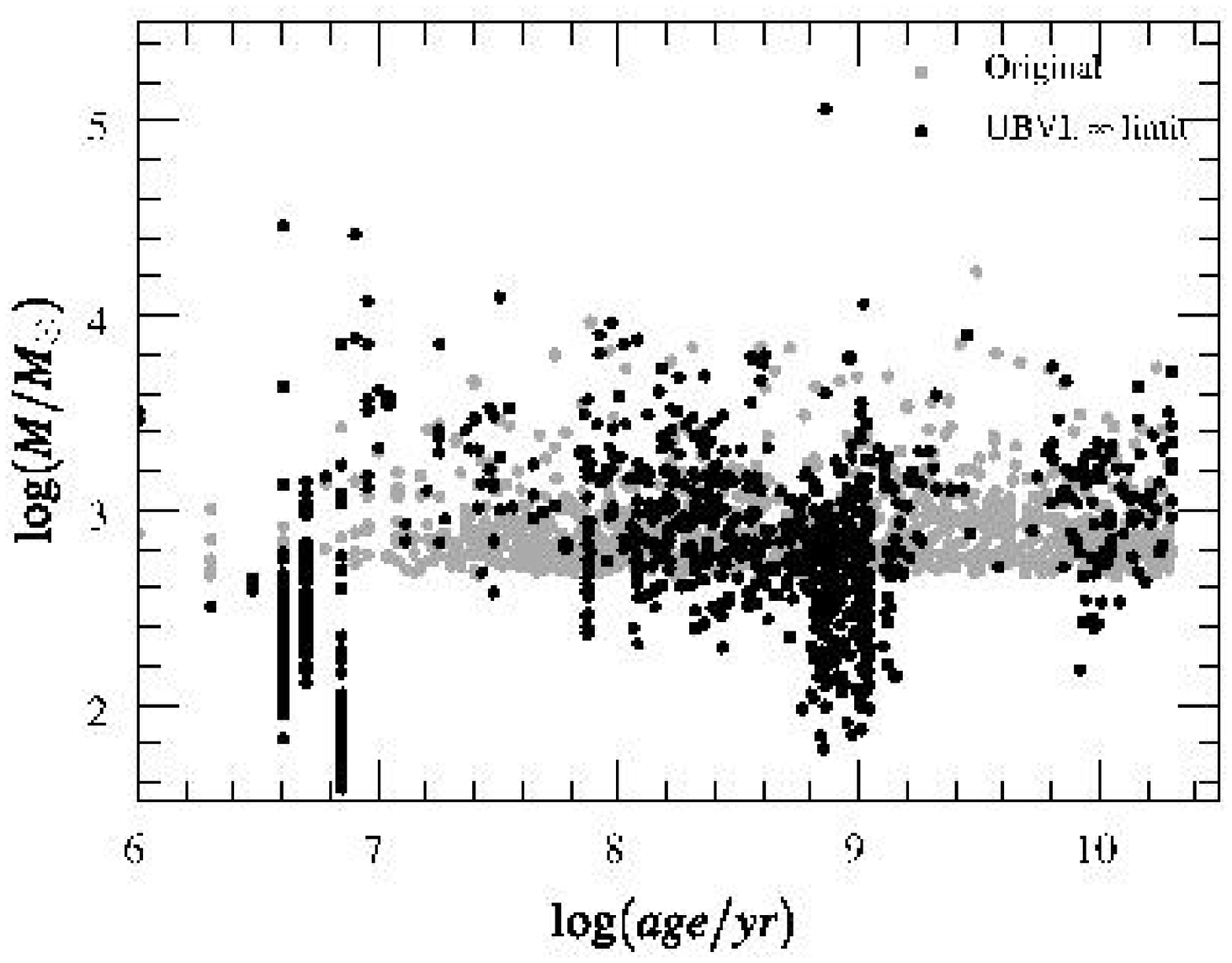}\hfill
        \includegraphics[width=8.8cm]{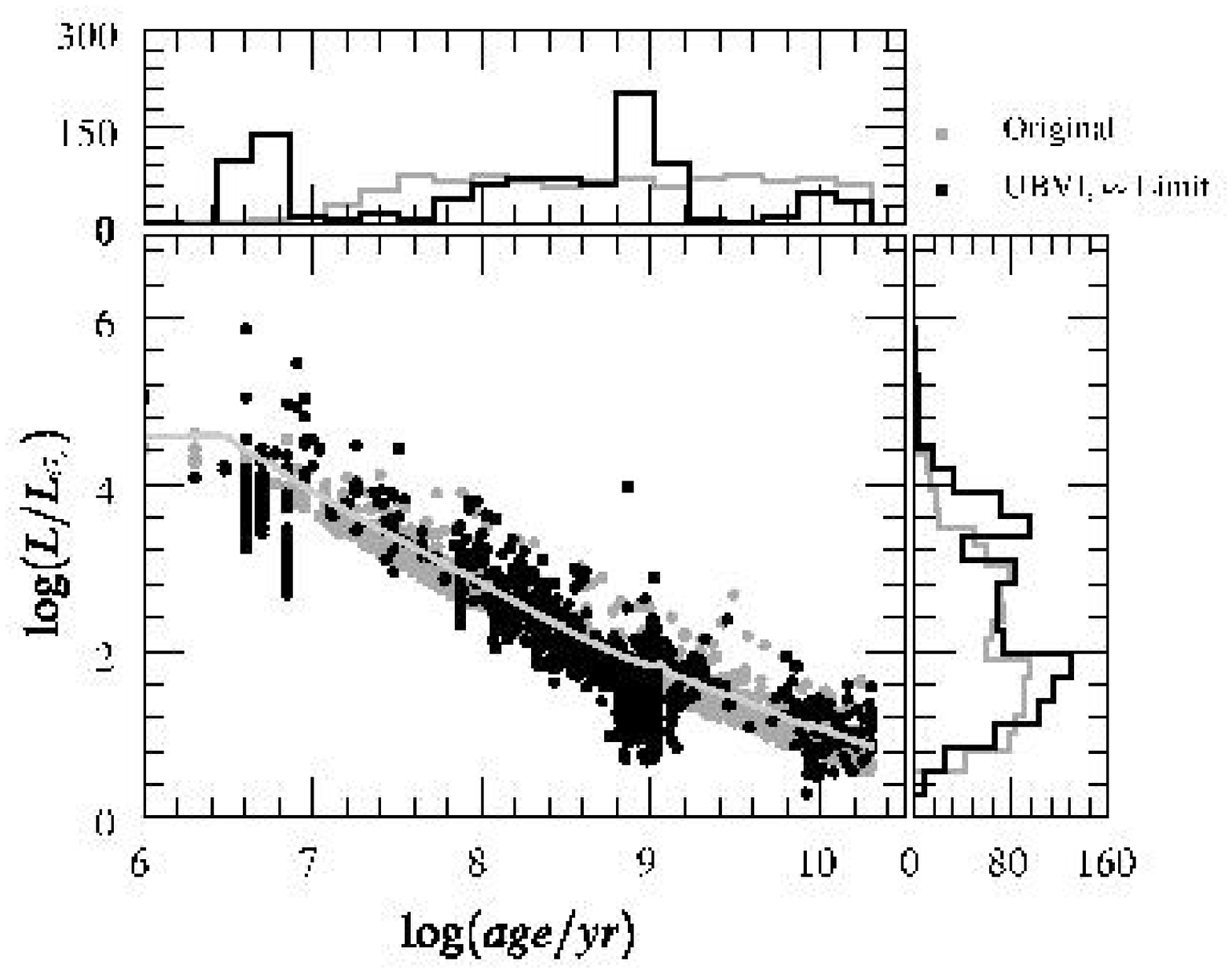}
	\caption{Age-mass and age-luminosity distributions obtained for our test
	sample, when analysing UBVI data with standard, continuous population
	synthesis models.  Grey dots represent the test-sample and black ones
	the recovered properties.  In the right panel, masses have been
	converted to luminosities using the (continuous) mass-to-light ratio
	given by P\'egase (solid line: $10^3$~M$_\odot$ cluster). }
	\label{fig:AgeMassLum_UBVI}
\end{figure*}

Figure~\ref{fig:AgeMassLum_UBVI} shows the age-mass distribution obtained when
UBVI data for our test sample of clusters is analysed with non-stochastic
population synthesis models and no fit is rejected.  While the properties of the
test sample are smoothly distributed, the estimated properties are highly
clustered. The figure looks similar with UBVIK data, but we chose UBVI for
comparison with the (UBVI+H$\alpha$)-based age-luminosity distribution of (more
massive) Antennae clusters \citep{Fall2005}. The right panel of
Fig.~\ref{fig:AgeMassLum_UBVI} shows the derived distribution in the
age-luminosity plane, after conversion of the estimated masses into luminosities
with the age-dependent mass-to-light ratio given by P\'egase.  In an
observational context, such a figure would be truncated at low luminosity by
some instrumental sensitivity limit.

The similarity between the distribution in Fig.~\ref{fig:AgeMassLum_UBVI} and
that in Fig.~$1$ of \cite{Fall2005} is striking, even more so when one mentally
corrects for the sharp low-mass limit of our sample and our relative lack of
young clusters. Accumulations and gaps occur at essentially the same ages
(differences can be traced back to the different sets of evolutionary tracks
used by the authors).  We have discussed the origins of the artefacts in our
test-sample age distribution in Sect.~\ref{sec:UBVIKphot}: they arise because
real, stochastic clusters of small masses are distributed over a wide range of
colours, and can therefore fall quite far off on the line of continuous models
when only extinction is available to alter their colours. The clusters in the
Antennae sample, however, are typically 50 times more luminous than the ones in
our study. More massive clusters should lie closer to the lines of continuous
models. Why then are these artefacts so strongly seen in the Antennae cluster
distribution? Our interpretation is that the combined effects of observational
errors and real extinction disperse the clusters enough in colour-space to
produce a global derived age distribution that is similar to the distribution
produced at lower masses by stochasticity.   

The input age distribution of the clusters in our test-sample (and in our main
MC catalog of clusters) is constant in logarithmic age bins, and this produces
derived age-luminosity distributions that are very similar to the ones in the
literature.  Our tentative conclusion is that the adopted age-distribution is,
for the time being, an adequate prior for the Bayesian, stochastic studies.
Clearly, it would be desirable to explore quantitatively how sensitive the
derived distributions are to the actual age and mass distributions, especially
in the presence of observational errors. Any firm conclusion on the Antennae
clusters, for instance, would require such a study.

\subsection{Prospects of the stochastic analysis}
\label{sec:prospects}

The analysis of cluster colours based on a library of stochastic models provides
ages and masses with small random errors, and with systematics that will be
negligible in many situations.  

\begin{figure*}
        \includegraphics[width=8.8cm]{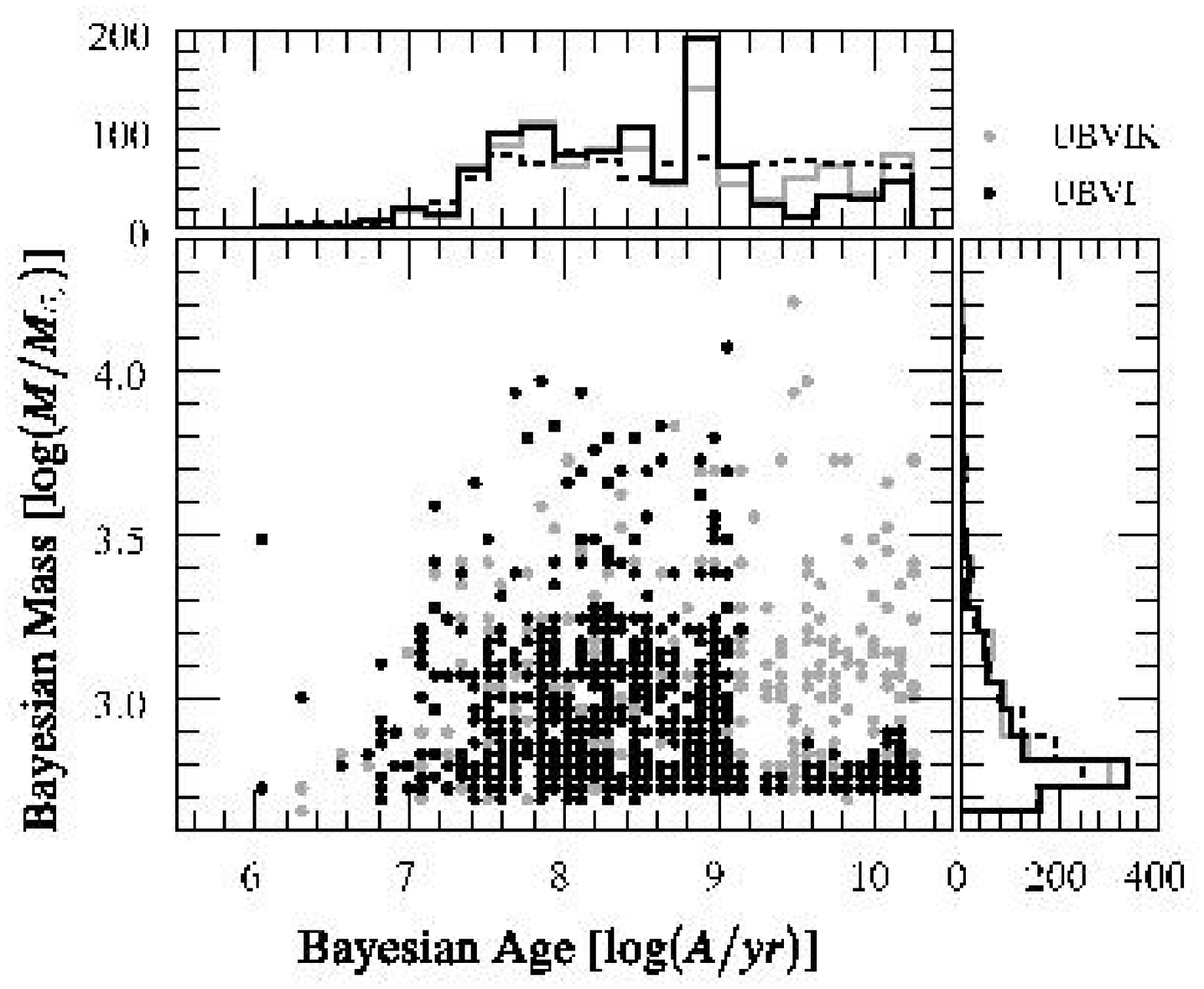}\hfill
        \includegraphics[width=8.8cm]{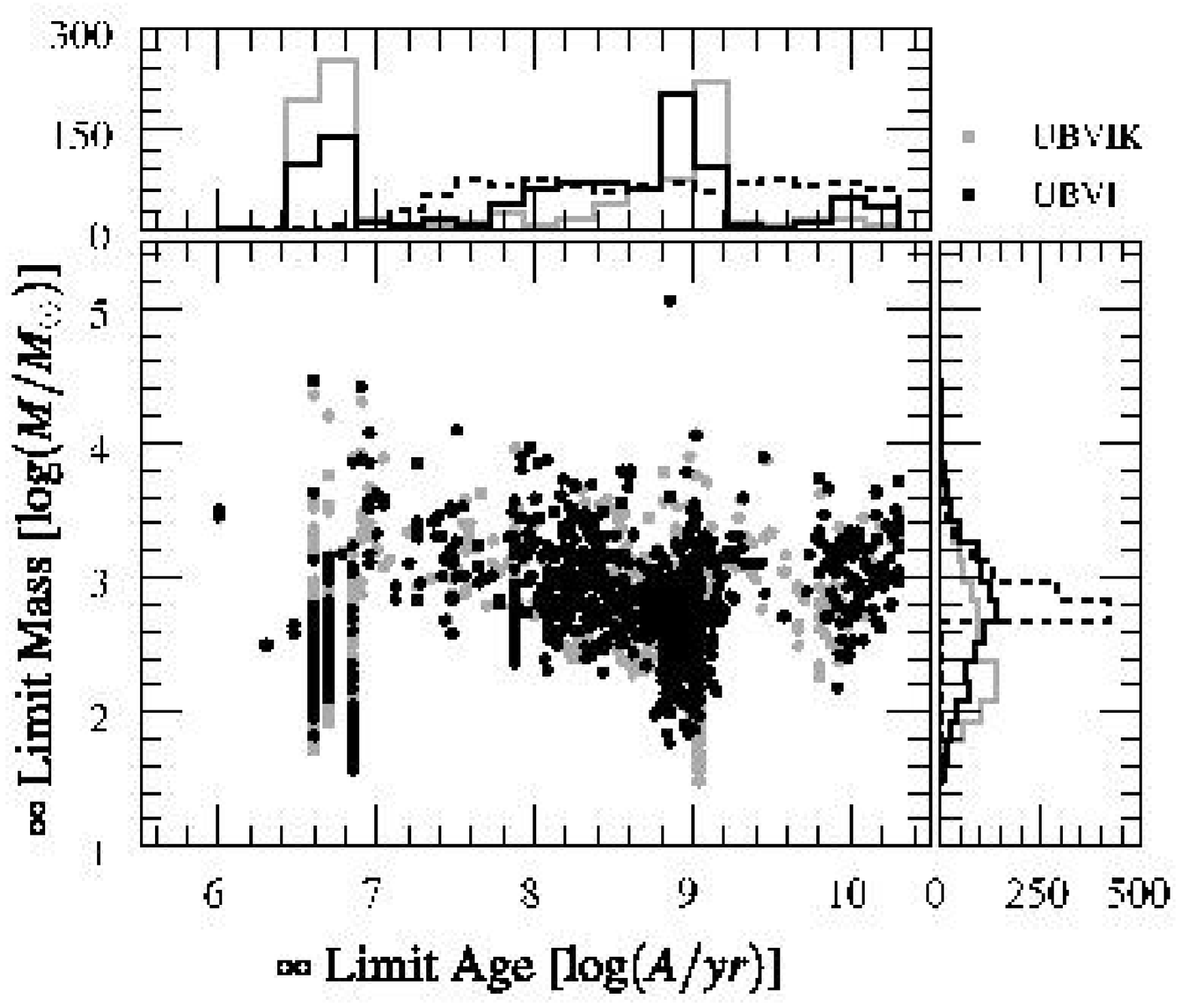}
	\caption{Resulting age-mass distributions from the Bayesian (left panel)
	and standard (right panel) analysis methods on the test-sample clusters.
	Grey dots are estimates from UBVIK bands whereas black dots are from
	UBVI bands only. UBVIK distribution on the right panel is presented in
	left panel of Fig.~\ref{fig:AgeMassLum_UBVI}.  Note that binning effects
	are visible on the Bayesian distribution as mentioned in the
	Sect.~\ref{sec:prospects}. The original age-mass distribution is
	presented on Fig.~\ref{fig:UBVIK_nonoise_sample}.
	\label{fig:UBVIK_distribs}
	}
\end{figure*}

\begin{figure}
        \includegraphics[width=8.8cm]{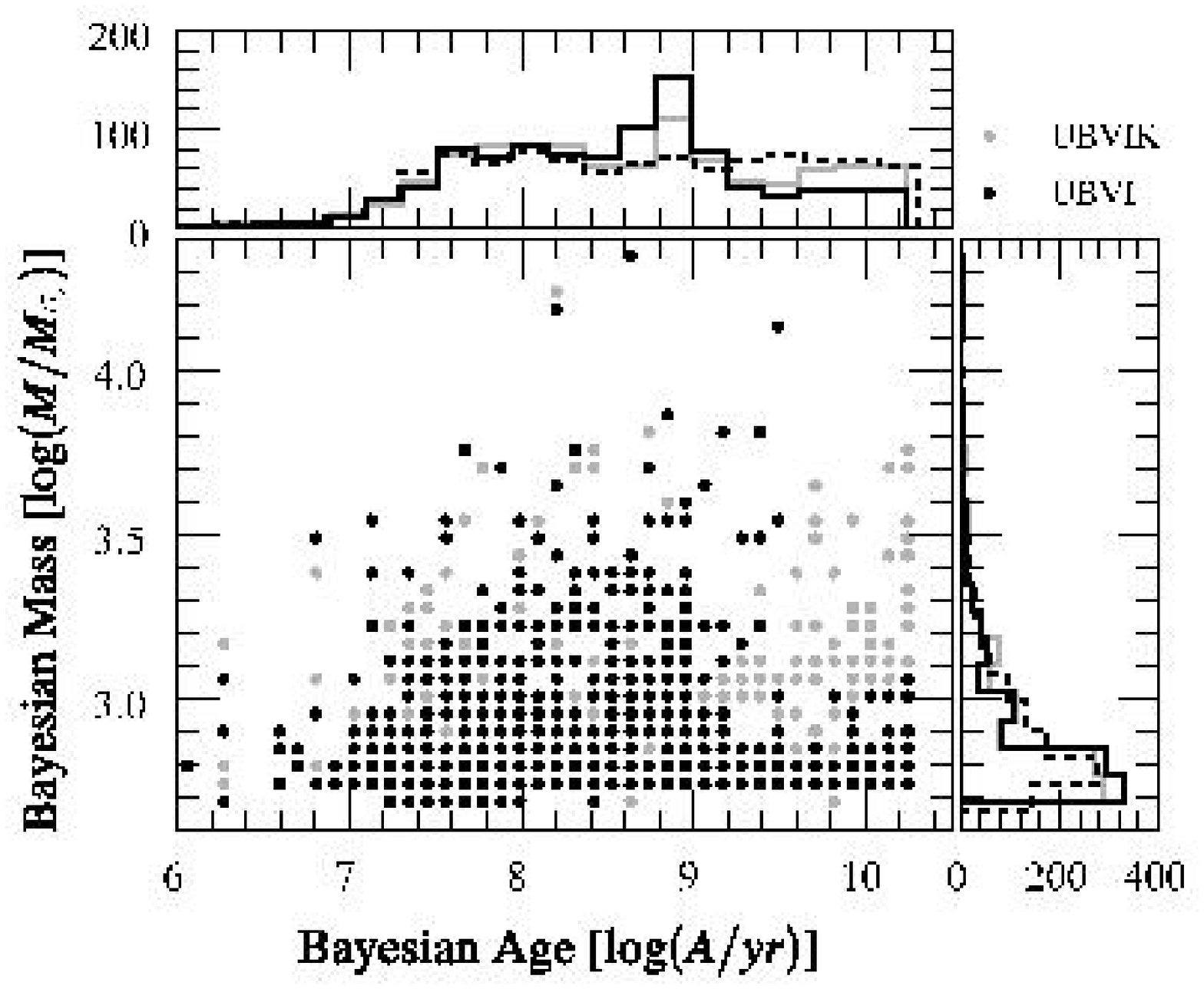}\hfill
	\caption{Resulting age-mass distributions from the Bayesian 
	analysis methods on the $1$-mag reddened test-sample clusters. 
	Grey dots are estimates from UBVIK bands whereas black dots are from
	UBVI bands only. Corresponding marginal
	distribution histograms follow the same color conventions. The dotted
	line on the histograms represent the test-sample distribution.
	\label{fig:UBVI_AV0_distribs}
	}
\end{figure}

Figure~\ref{fig:UBVIK_distribs} shows the derived age-mass distributions for our
test sample, based on either UBVI or UBVIK data sets. {Similar results are
obtained if the input sample is reddened (Fig.~\ref{fig:UBVI_AV0_distribs}).}
The situation can be improved even more if there are independent constraints on
extinction.

The figures show the grid pattern due to the finite sizes of our bins in
estimated ages and masses. This is because we compute the posterior
probabilities of Eq.~\ref{eq:proba} for finite intervals. 

The binning procedure has a caveat and an advantage.  The caveat is that results
are sometimes sensitive to the choice of the bin sizes or boundaries. We have
looked at many probability maps in 2D projections of age-mass-extinction space
to check where this occurs.  Examples are given in Appendix \ref{sec:Apxmaps}
(all can not be developed here).  As expected, the results are robust with
respect to the binning (i.e. errors are smaller than or equal to one bin size)
when the maps are simple, single peaked (Fig.~\ref{fig:UBVIK_probmap}), while
they can depend strongly on binning when the maps are complex (i.e. in cases in
which uncertainty estimates based on contours of equal probability would give
large error bars).  The second type of situation occurs especially when there is
no {\em a priori} information on extinction.  Figure~\ref{fig:ExtinctionEffect}
illustrates this event: the dereddening vector of an object crosses several
regions of high model-density. The model properties within each of these regions
end up with similar probabilities.  Which one dominates, depends on the binning
as shown on Fig.~\ref{fig:BinningEffect} (and on the density pattern itself).

The advantage of binning is that we can, to first order,  assign the ages,
metallicities and extinction of a given bin the same {\em a priori}
probabilities. This allows us to rescale the derived probability maps for a
cluster after calculation, if we wish to change the adopted priors (a worthwhile
gain in computation time).  In our current catalog, the total number of clusters
with masses above about $10^4$~M$_{\odot}$ is too small to allow us to study
flatter mass distributions of clusters immediately, and we also lack clusters
with ages below $10^7$~yr.  We have therefore chosen to postpone the description
of consequences of changes in the priors to a future article. 

\begin{figure*}
        \includegraphics[width=9.2cm]{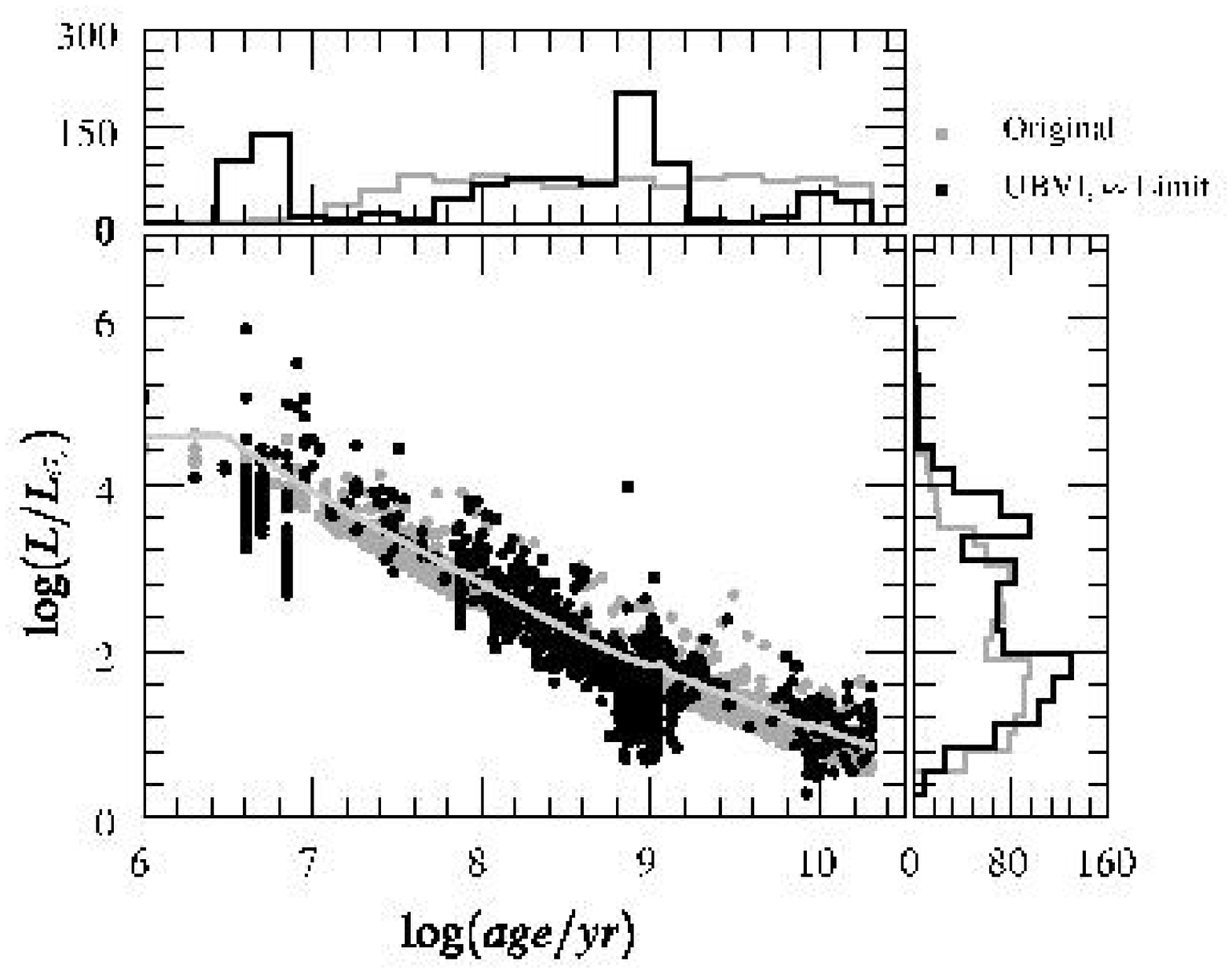}\hfill
        \includegraphics[width=9.2cm]{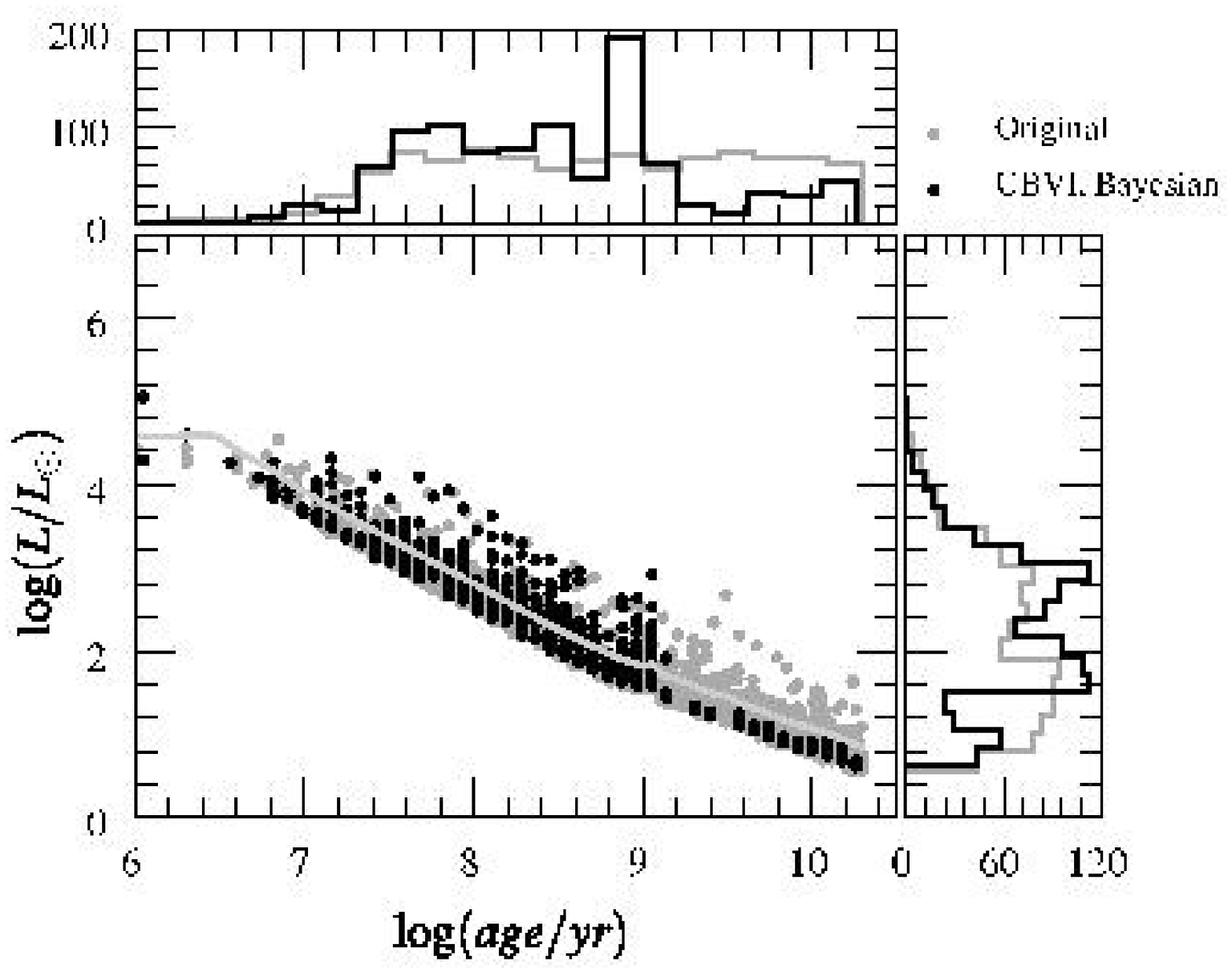}
        \includegraphics[width=9.2cm]{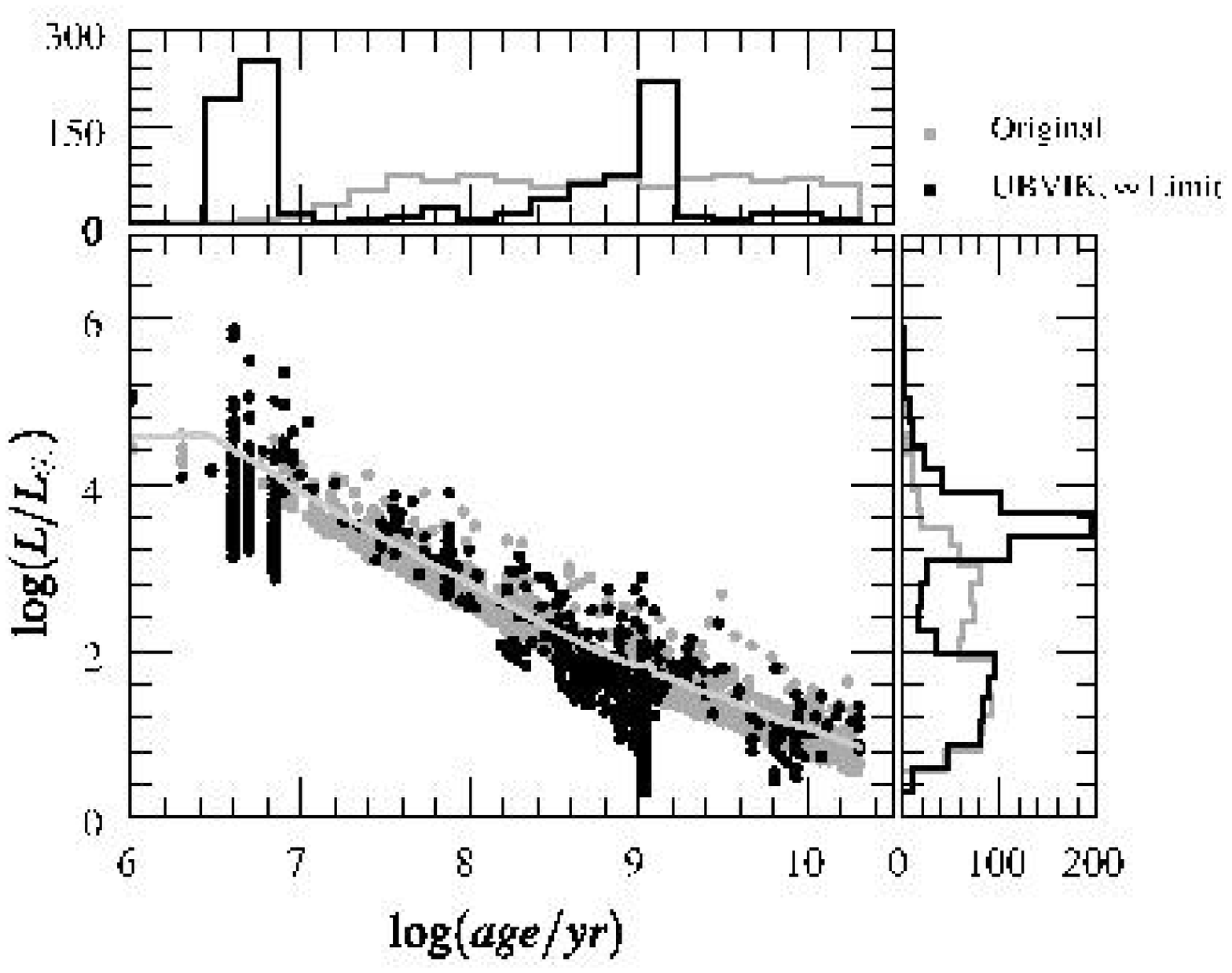}\hfill
        \includegraphics[width=9.2cm]{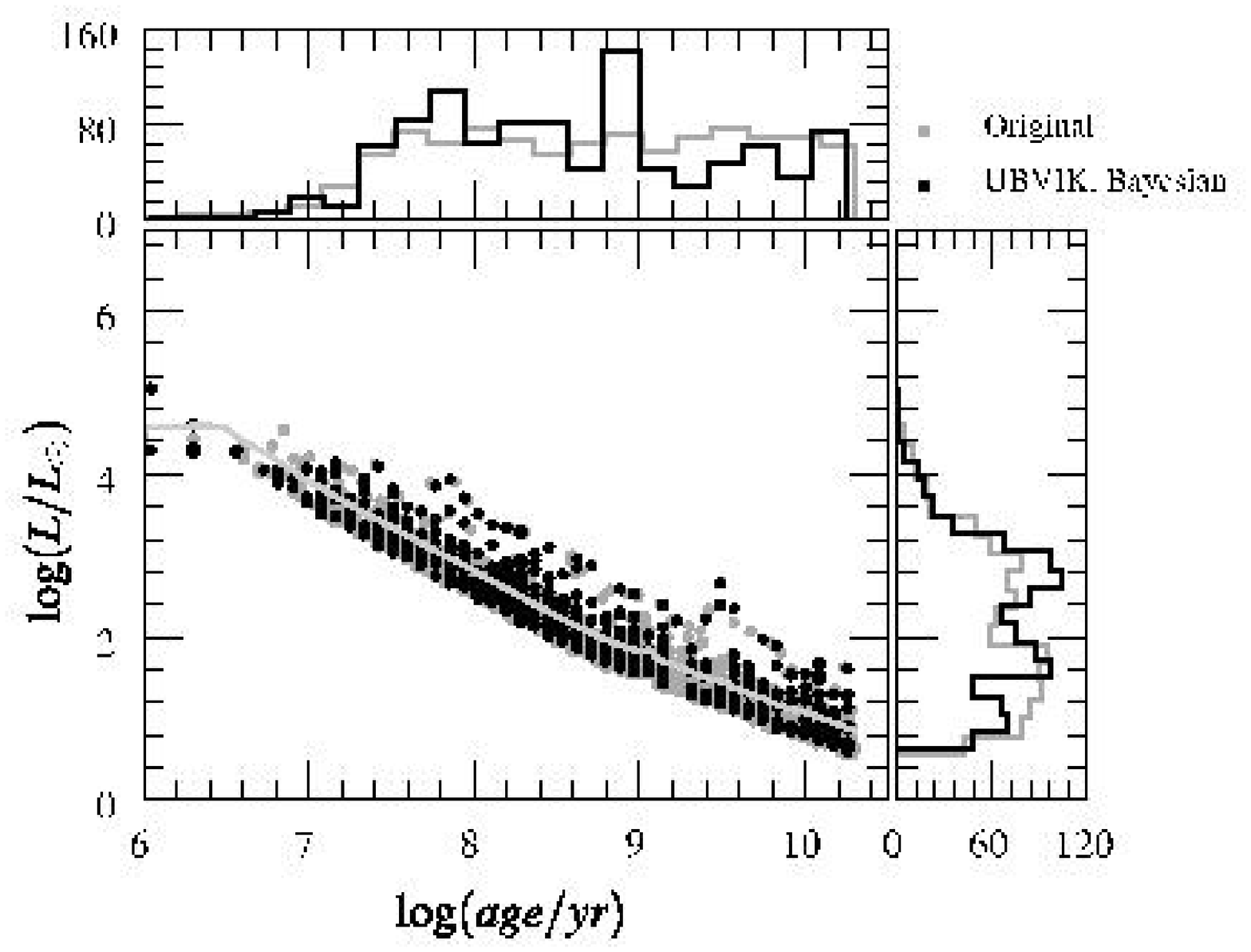}
	\caption{Age--luminosity recovered distributions obtained for our
	test-sample.  Grey dots are the original distributions and black dots
	are respectively the recovered distributions from the standard (left
	panels) and Bayesian (right panels) methods when analysing UBVI (top
	panels) and UBVIK (bottom panels) data sets. Recovered masses are
	converted to luminosities using P\'egase (continuous) factor of which
	the solid lines represent the evolution of a $10^3$~M$_\odot$ cluster.
	\label{fig:Age_Lum_Distribs}
	}
\end{figure*}

\subsubsection{Other practical aspects}
\label{sec:cpu}
Although conceptually simple, the implementation of the Bayesian, stochastic
analysis is not immediate. Constructing the stochastic library requires CPU time
and data storage. For each cluster, the spectra of at least the brightest 1000
stars need to be summed individually. The remaining low mass stars could be
represented with a continuous sequence to save computation time.  Synthetic
broad band colours can then be measured on the resulting total spectrum. We
decided to save all the spectra and to compute a set of 60 standard broad band
fluxes  (standard UBVRIJHK, but also most HST filters and some from other
telescopes). For three metallicities and several millions of models in total,
the computation took a couple of month on six $3$~GHz CPUs, and half a
terabyte of data was accumulated. 

Browsing through the collection to compute probabilities also takes time, even
with the restriction to a few photometric bands. A few  CPU-hours are required
to construct probability distributions of age, mass and extinction for $1000$
clusters with our main MC catalog, allowing for extinction to vary in the range
described above. It is somewhat faster to use a simple best-$\chi^2$. 
This may be
good enough for statistical studies of large samples, but it must be kept in
mind that the single best fit can be far from the Bayesian most probable model in
individual cases. {Very small changes in the observed colours (due to small
observational errors) can modify single best fit parameters, while they
leave Bayesian estimates unchanged.
To provide error bars on the single best fit parameters, one has to
explore the vicinity of the minimum $\chi^2$. For instance, one could
compare the properties of all the models with a 
$\chi^2$ below a predefined threshold, as was done for large clusters
in \citet{Bridzius2008} or \citet{Anders2004}. Doing this exhaustively 
would be conceptually similar, and essentially as long to run, as
our Bayesian approach.}

All the results in this paper are based on one single set of population
synthesis models. {Some caveats of these models are that they do not include the
formation of carbon stars or Mira-type variability, and that they do not account
for binaries or for stellar rotation}. We have not yet compared the amplitude of
the errors arising from stochasticity (studied here) with those due to
uncertainties in the physical assumptions of the models.  However, we expect
qualitatively similar systematics with any set of stellar evolution tracks.  It
will unfortunately be necessary to recompute a completely new library if one
wishes to explore different sets of stellar evolution tracks, to assume a
different stellar mass function, or to use a different library of stellar
spectra.  This will clearly limit the pace at which the comparisons can be made.

\subsection{Impact of the photometric band selections}
\label{sec:bandselection}

In the present paper, the estimation of ages and masses has been shown to be
affected by stochasticity and also by the photometric datasets used during the
analysis. We chose to focus on explaining behaviours rather than multiplying
experiments with different pass-bands, and have therefore restricted our
quantitative comparisons to two combinations. In a large fraction of stellar
population studies, the UBVI combination is used as the standard selection: it
provides a reasonable compromise between depth and spectral coverage for most of
the available instruments. Moreover, it corresponds to a wavelength range where
spectral libraries are the most accurate. Adding the K band improves mass
estimates in the non-stochastic regime of high total masses, but it has an
observational cost as it requires the use of a different instrument. Testing the
UBVIK combination in the stochastic regime provides useful information for the
design of future surveys.

In Sect.~\ref{sec:noKband}, we presented some effects of the presence of K band
data in addition to UBVI photometry. The K band does not improve significantly
the situation when {\em continuous} models are being used (right panel on
Fig.~\ref{fig:UBVIK_distribs}).  In particular, we mentioned that including the
K band in the analysis of populations with AGB stars ($100-500$ Myr) may lead to
even worse estimates (again with continuous models). This effect is also visible
on the left-hand side panels of Fig.~\ref{fig:Age_Lum_Distribs}.  On the other
hand, we found only mild improvements when including K band data in the {\em
Bayesian} analysis: the presence of K band information translates into weaker
artefacts in the derived age distributions (artefacts which disappear when
reddening is not a free parameter).  The left panel of
Fig.~\ref{fig:UBVIK_distribs} presents recovered age-mass distributions for UBVI
and UBVIK datasets using the Bayesian approach, with free extinction.  Old
populations ($> 1$~Gyr) age-mass estimates reflect the original distribution
better when K band constraints are included.  This is also visible on the right
panels of Fig.~\ref{fig:Age_Lum_Distribs}, where the resulting distributions are
significantly improved. 

At young ages, both our test sample and our main MC catalog are underpopulated.
Most small and young clusters have HR diagrams that resemble a truncated main
sequence. Even if the stellar mass function from which their stars are drawn
randomly extends to 120\,M$_{\odot}$, a small zero age cluster will in general
contain zero ionizing stars, and a small cluster aged 10\,Myr will contain zero
red supergiants : two such objects are basically indistinguishable, be it with
or without K band data.  This problem shows up clearly when we run the Bayesian
analysis at young ages, but quantifying this effect requires that we add more
young clusters to our reference catalog. 

\smallskip

The UV spectral range raises questions similar to the near-infrared.  We
repeated some of the above experiments with UBVI and the F218W filter, centered
around $220$\,nm (WFPC2 instrument on-board the Hubble Space Telescope).  A brief
summary is that the effects of adding this UV band have amplitudes similar to
those obtained when adding K.  The dispersions in the results are comparable to
those without UV, both for Bayesian or standard estimates.  Artefacts are also
qualitatively similar. Age estimates below $20$~Myr are improved only slightly.

If one however wants to include UV bands, the choice of an extinction law
becomes more of an issue.  Extinction laws indeed express a range of behaviours
in the UV bands \citep[e.g][]{Allen1976, Fitzpatrick1986, Cardelli1989,
Calzetti2000}.

The decision of which filters to use rests on considerations of the
astrophysical problem to tackle.  Using either UV or K photometry does not
deeply affect the resulting estimates ages and masses of small clusters.  One
may first want to consider using discrete population models instead of
continuous ones.

\section{Conclusions} 
Studies of star cluster populations in galaxies have been based until now on
{\em continuous} population synthesis models, that provide a very poor
approximation of the integrated light of clusters of small and intermediate
masses because this light is determined by a very small number of luminous
stars.

Based on large collections of Monte-Carlo simulations of star clusters that each
contain a finite number of stars, this paper explores systematic errors that
occur when the integrated fluxes of realistic clusters of small masses are
analysed in terms of mass and age. Our main collection is built with the cluster
age and mass distributions of \citet{Fall2009}, extrapolated to masses lower
than those observed in the Antennae galaxies.

With the standard methods (continuous models), large systematic errors affect
estimated ages and large random errors affect masses.  If observational
uncertainties on cluster fluxes are large and, as a consequence, quality-of-fit
criteria fail to reject the numerous poor fits, systematic errors (of a few
tenths of a dex) are also present in the estimated masses.  Derived age-mass or
age-luminosity distributions for samples in which actual ages and masses are
distributed as in our main collection display clustered patterns that very
closely resemble those found in empirical samples in the current literature. We
find no immediately obvious reason to reject the age and mass distributions of
our main collection, but clearly this essential point requires detailed study
with real observations. 

A Bayesian method has been described and implemented, in order to account
explicitly for the finite nature of clusters in the analysis. It is shown that
their age and mass can be recovered with error bars that will be small enough
for many purposes. Young small mass clusters will remain difficult to age-date
because HR-diagrams of many of them identically look like truncated main
sequences with no ionizing or post-main sequence stars.  At intermediate ages,
the variability of luminous AGB stars is expected to cause difficulties that we
have not yet solved.

The comparison between the results obtained with UBVI and UBVIK data sets shows
that, in the stochastic context, the benefits of adding the K band to optical
observations are rather small, except for the mass determination of clusters
older than 1\,Gyr.  Clearly, adding near-IR or UV information is secondary,
compared to the need to move from continuous to stochastic cluster models.

The Bayesian analysis method can now be applied to existing data on cluster
samples in nearby galaxies with the aim of constraining the actual age and mass
distributions of these clusters. We will also extend the study of systematic
errors to the case where metallicity is an unknown parameter.

\begin{acknowledgements}
We are grateful to Michel Fioc for kindly providing his Fortran~90 version of
{\sc Pegase}, and to M. Fioc and B. Rocca-Volmerange for their support to the
implementation of discrete population synthesis.
\end{acknowledgements}

\bibliographystyle{aa}

\appendix 

\section{Examples of probability maps}
\label{sec:Apxmaps}

The method we developed to take the stochasticity into account to estimate
intrinsic parameters (age, total number of stars or total mass, metallicity,
extinction) of unresolved population follows a Bayesian approach. Given a set of
photometric measurements and uncertainties, we establish the joint
probability distributions of these parameters based on a large catalog of
Monte-Carlo simulations.  
In this appendix, we present examples of probability distributions.

\begin{figure*}
        \includegraphics[width=8.8cm]{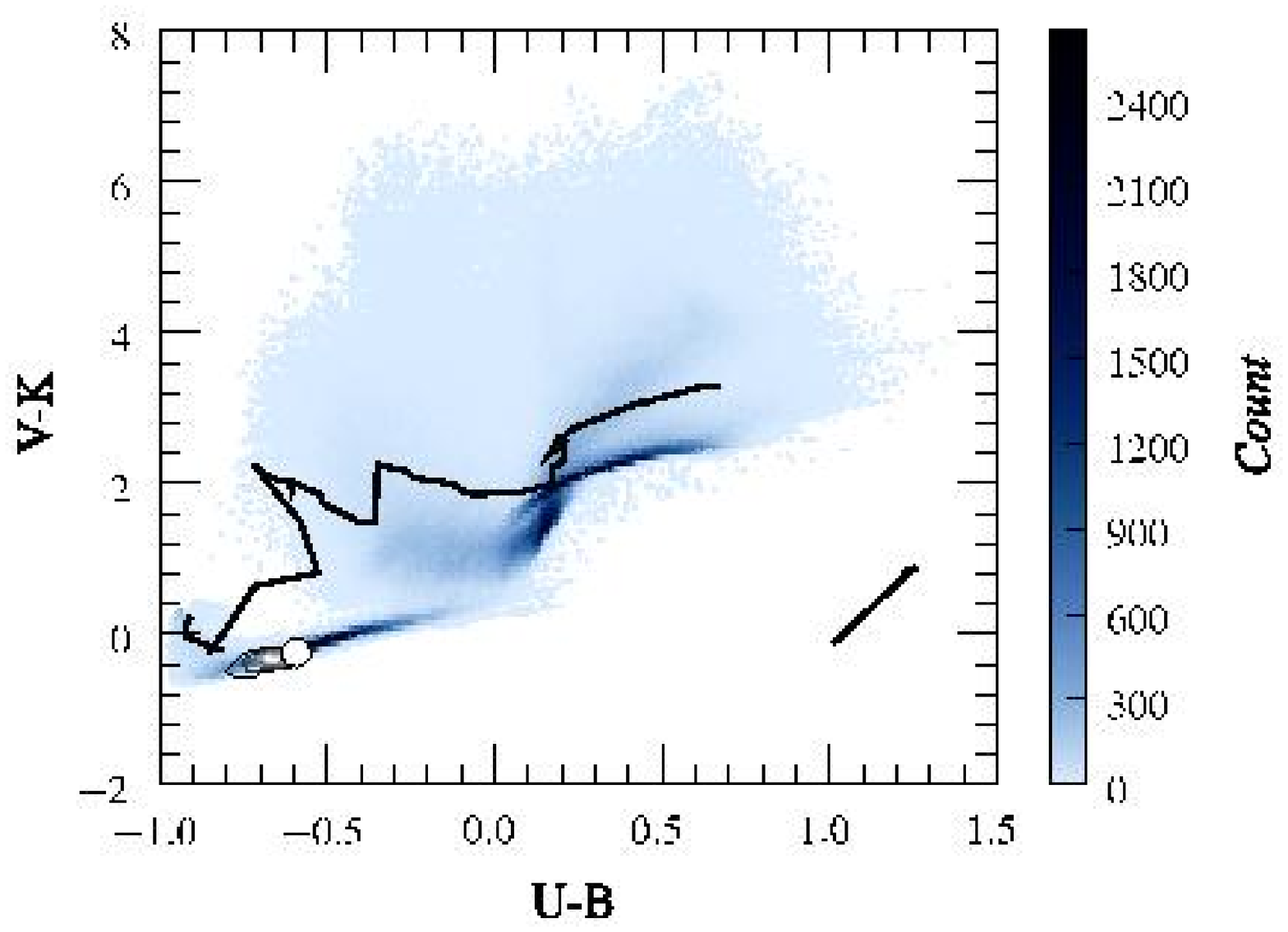}\hfill
        \includegraphics[width=8.8cm]{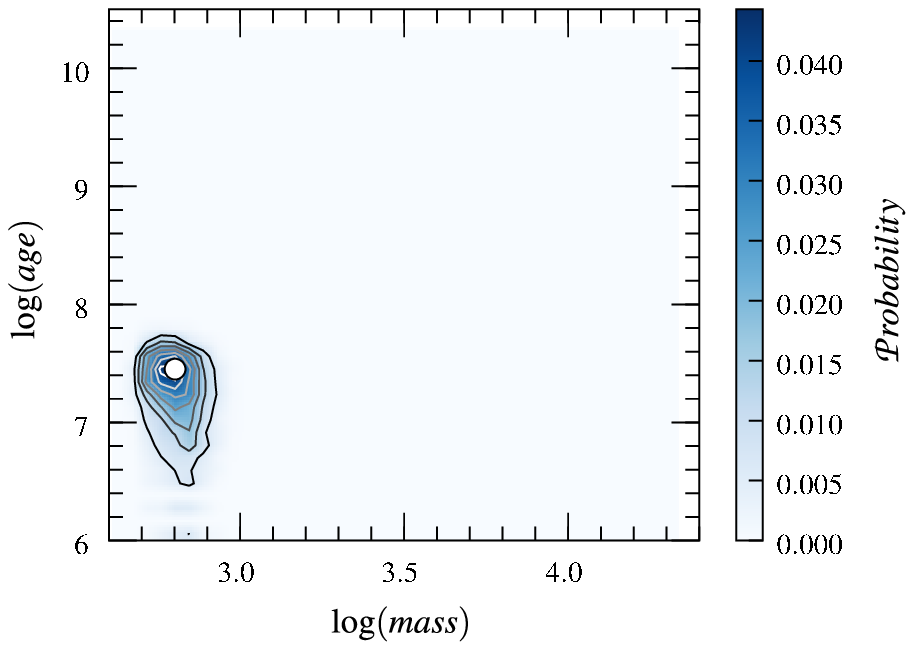}\\
        \includegraphics[width=8.8cm]{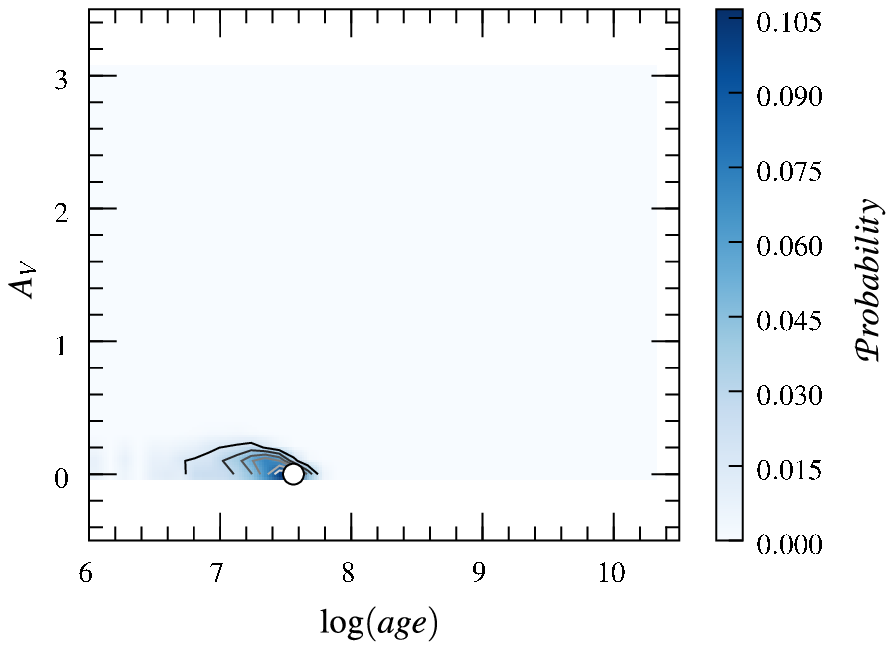}\hfill
        \includegraphics[width=8.8cm]{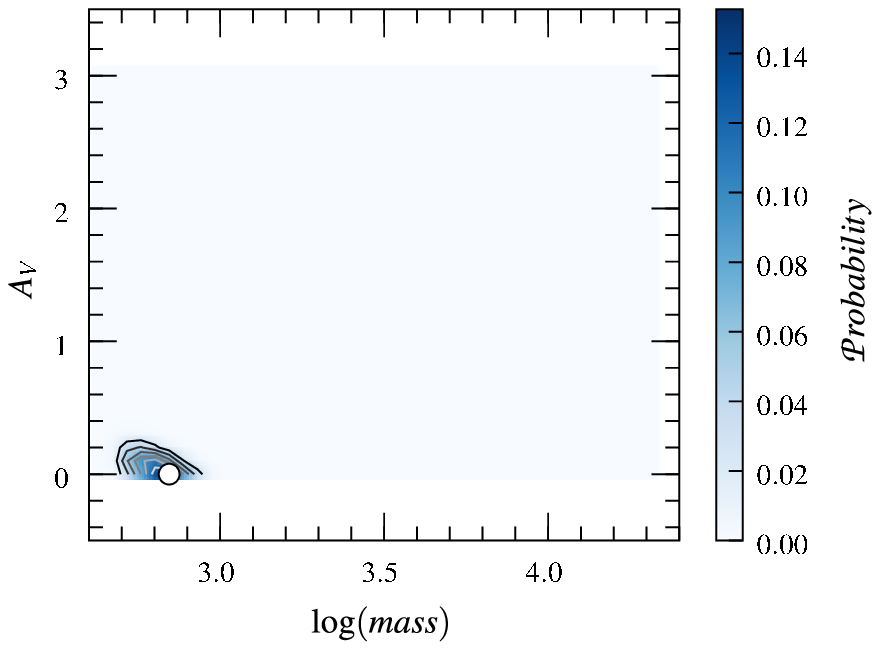}\\
	
	\caption{Example of probability map for a population of $33$~Myr and
	$650$~M$_{\odot}$ based on UBVIK band data, allowing extinction to vary
	between $A_V=0$ and $A_V=3$. On the top-left panel, contours use an
	arbitrary scale to represent the colour joint probability distribution
	derived from the measurements (i.e.  Eq.~\ref{eq:proba} applied to
	colours); the colour scale refers to model density. The white circle
	indicates the locus of the studied population (error bars of $0.05$ mag
	are not indicated).  On the other panels, the colour scale refers to the
	corresponding 2D probability distributions. They are underlined by
	contours with arbitrary levels.  The white circles indicate the position
	of the highest probability.
	\label{fig:UBVIK_probmap}
	}
\end{figure*}

First an example is given on Fig.~\ref{fig:UBVIK_probmap}, for which we
obtain single peaked probability distributions leading to an unambiguous
probabilistic determination of age, mass and extinction estimates. 
In this example, we recover that the most likely have no extinction and 
the most probable age and mass are $35$~Myr and $630$~M$_\odot$, very close to
the expected values.

\begin{figure*}
        \includegraphics[width=8.8cm]{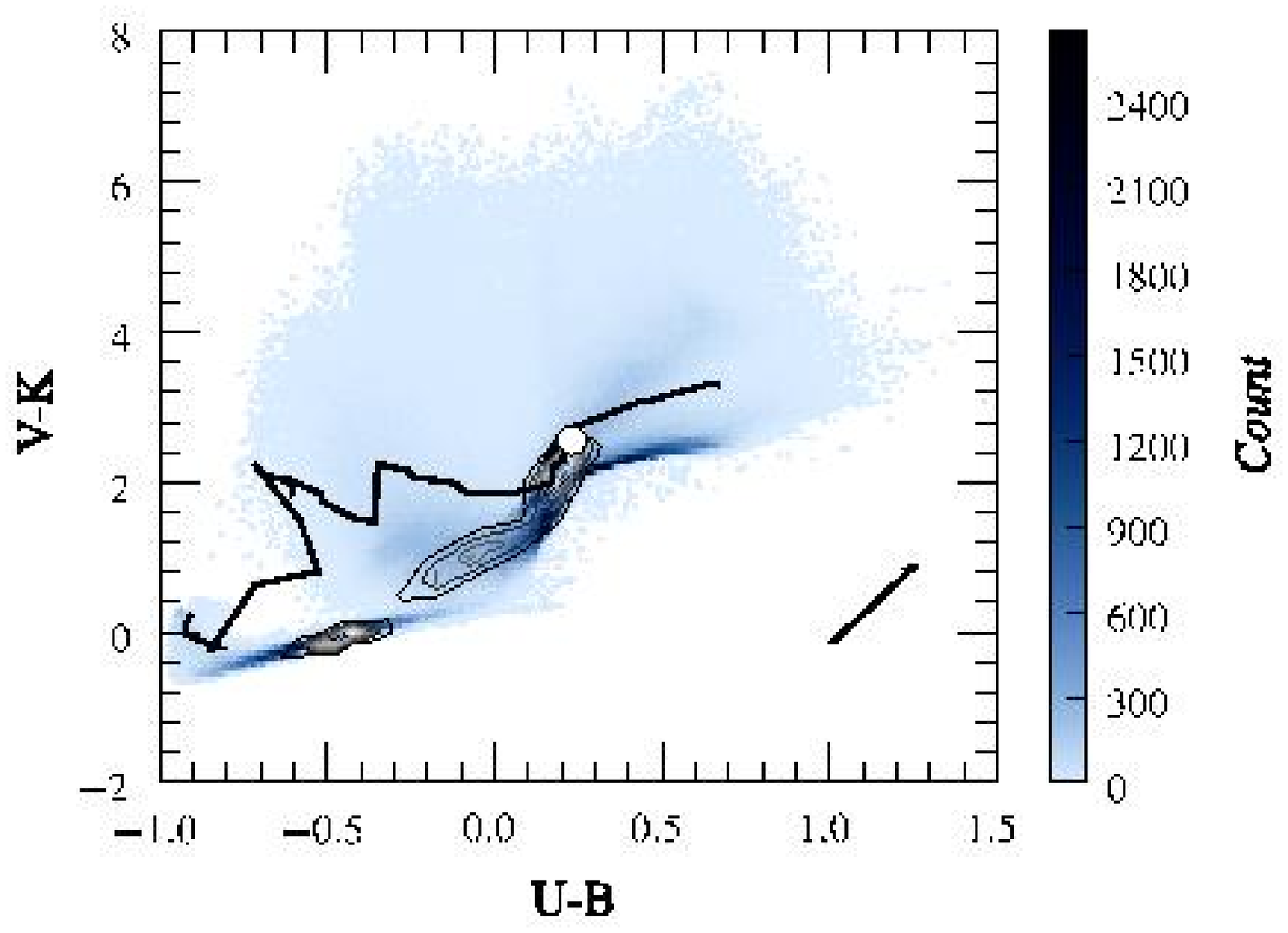}\hfill
        \includegraphics[width=8.8cm]{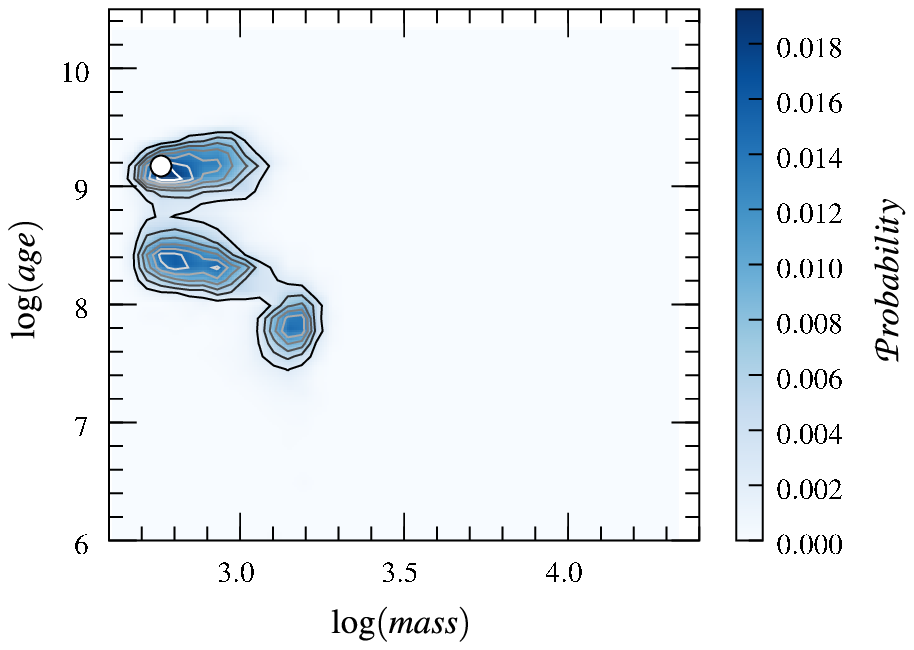}\\
        \includegraphics[width=8.8cm]{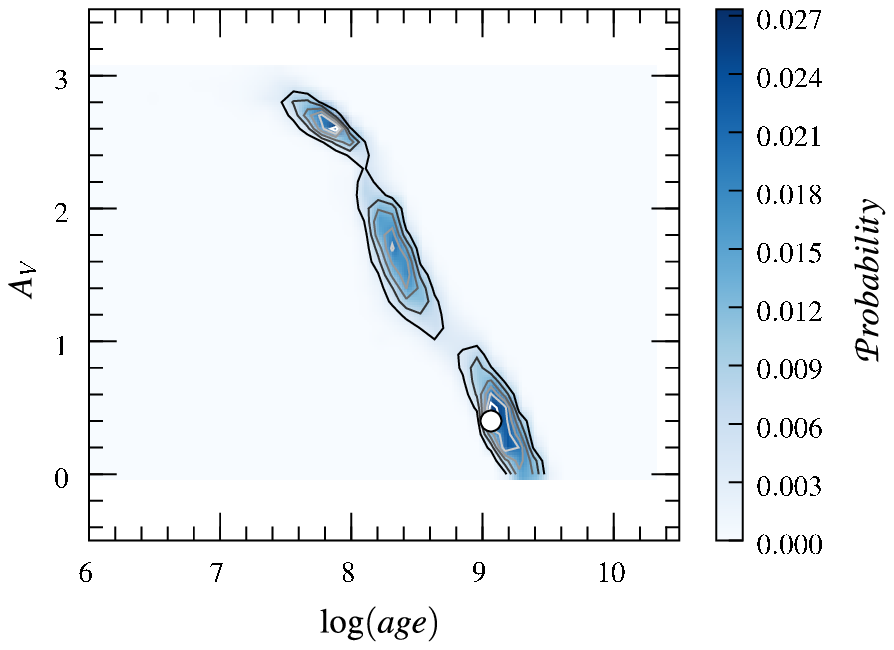}\hfill
        \includegraphics[width=8.8cm]{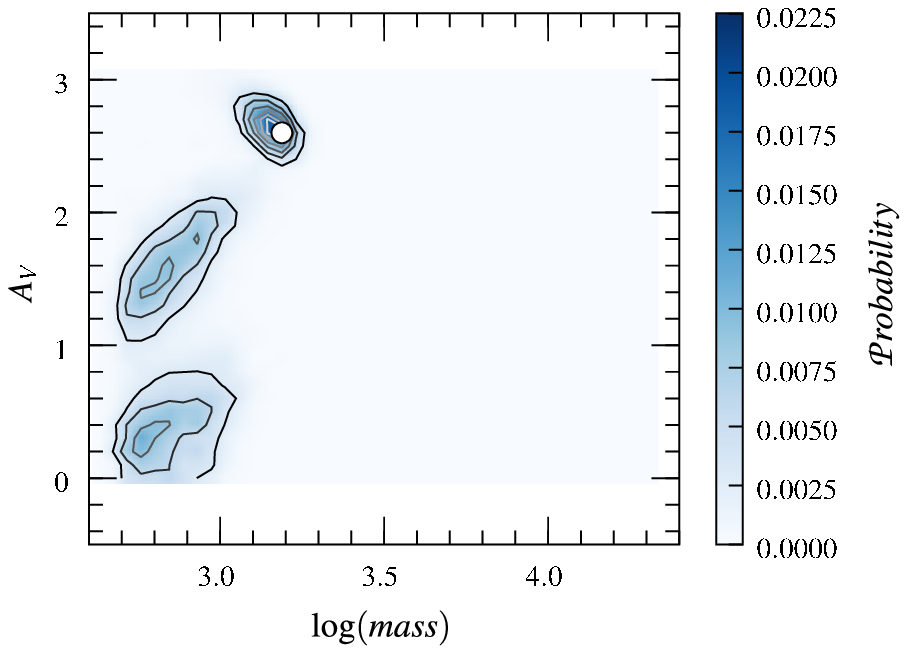}\\
	
	\caption{Example of probability maps for a population of
	$2.3$~Gyr and $780$~M$_{\odot}$ based on UBVIK band.
	Symbols and contours are the same as on Fig.~\ref{fig:UBVIK_probmap}.
	Age and mass estimates have been computed over 40 bins (refer to
	the binning issue described in Sect.~\ref{sec:prospects} and also related to
	Fig.~\ref{fig:BinningEffect}).
	\label{fig:ExtinctionEffect}
	}
\end{figure*}

As we discussed in Sect.~\ref{sec:prospects}, it also happens that the probability
distributions are not single peaked. Complex maps occur when there is no
independent known information on the amount of extinction.
Figure~\ref{fig:ExtinctionEffect} illustrates how the extinction factor 
increases the complexity of the probability maps. 
The top-left panel contours show how
the color distributions are affected along the dereddening vector. On this
example, the object crosses three regions of high model density. The resulting
age-mass probability distribution is multi-modal (trimodal). 

\begin{figure*}
        \includegraphics[width=8.8cm]{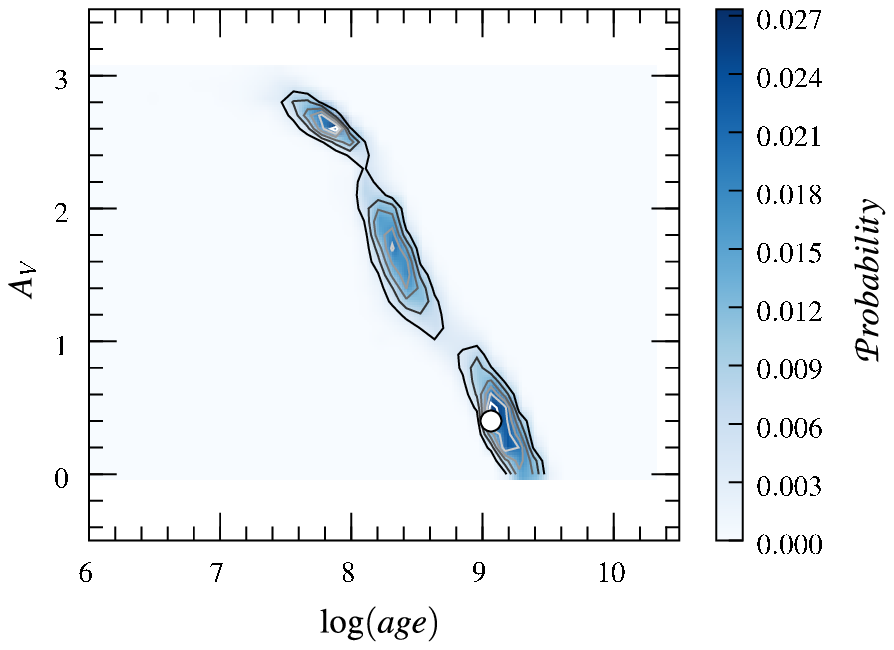}\hfill
        \includegraphics[width=8.8cm]{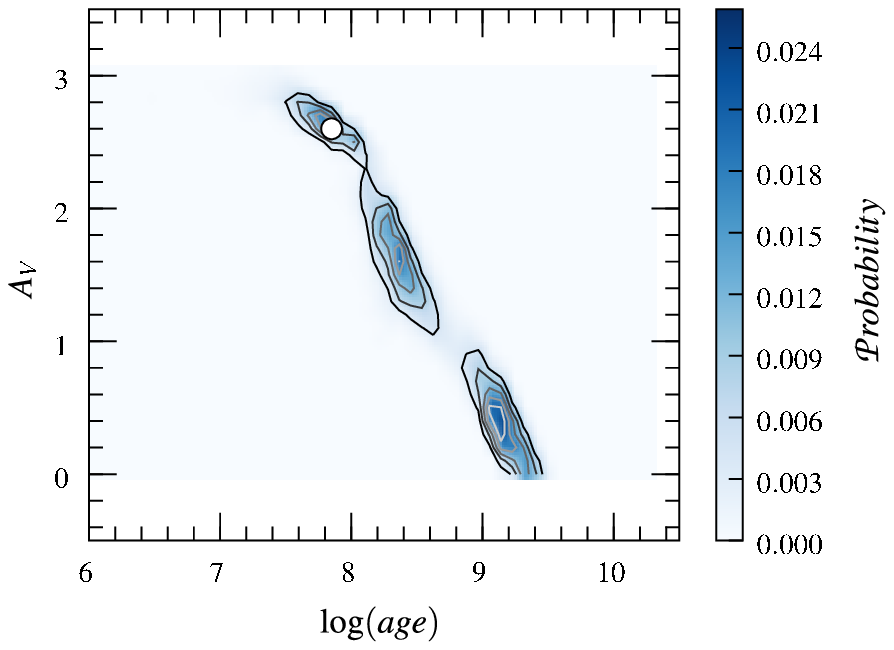}
	
	\caption{Illustration of the effect of binning using the cluster of
	Fig.~\ref{fig:ExtinctionEffect}. On left-hand side
	panel the correct age is recovered with 40 bins, however as the right 
	panel illustrates, a different binning (50 bins) favours a much younger
	age with (erroneously) high extinction. The cluster is now moved to
	feature \ref{fig:UBVIK_nonoise}$.b$ in Fig.\ref{fig:UBVIK_nonoise}.
	}
	\label{fig:BinningEffect}
\end{figure*}

Complex distributions are subject to binning issues. The example probability
distributions given on Fig.~\ref{fig:ExtinctionEffect} yields three modes with
similar probabilities. If one changes the number of bins on which the
probabilities are computed, peak values might vary to eventually change the
dominant mode of the distributions. The Figure~\ref{fig:BinningEffect} shows
that the variations of the resulting age estimates can be significant: from
$2.3$~Gyr to $50$~Myr .  A determination of the optimal binning remains an open
issue. 


\end{document}